\documentclass[useAMS,usenatbib, onecolumn]{mn2e}
\usepackage{fullpage}
\usepackage{amsmath}
\usepackage{amssymb}
\usepackage[a4paper, lmargin=0.9in,  rmargin=0.9in, 
tmargin = 0.8in, bmargin=0.8in, headsep=0.1in]{geometry}
\usepackage[utf8]{inputenc}
\usepackage[english]{babel}
\usepackage{enumitem}
\usepackage[dvipsnames]{xcolor}
\usepackage{graphicx}
\usepackage{graphics}
\usepackage{wrapfig}
\usepackage{hyperref}
\hypersetup{
     colorlinks   = true,
     citecolor    = blue}

\newcommand{\myemail}{wenbinlu@caltech.edu}

\def\apj{{ ApJ}}
\def\apjl{{ApJL}}

\def\aap{{ A\&A}}

\def\mnras{{ MNRAS}}
\def\araa{{ ARA\&A}}

\def\nat {{ Nature}}

\def\pasp{{PASP}}

\def\physrep{{ Physics Reports}}

\def\prl{{ Phys. Rev. Lett}}

\def\mr{\mathrm}
\def\mc{\mathcal}
\def\d{\mr{d}}

\def\aum{a_{\rm \mu m}}
\def\um{\mr{\mu m}}
\def\tmu{\tilde{\mu}}
\def\ttheta{\tilde{\theta}}
\def\tphi{\tilde{\phi}}
\def\trho{\tilde{\rho}}
\def\tx{\tilde{x}}
\def\ty{\tilde{y}}
\def\tz{\tilde{z}}
\def\QabsP{\langle Q_{\rm abs}\rangle_{\rm P}}

\def\gm{\gamma_{\rm m}}

\def\num{\nu_{\rm m}}
\def\epse{\epsilon_{\rm e}}
\def\epsB{\epsilon_{\rm B}}

\title[Infrared Dust Echo]{Infrared dust echoes from neutron star mergers}
\author[Lu, McKee \& Mooley]
  {Wenbin Lu$^{1, 2}$\thanks{\myemail}, Christopher F. McKee$^3$, and Kunal P. Mooley$^{4, 5}$\\
  $^1$Department of Astrophysical Sciences, Princeton University, Princeton, NJ 08544, USA\\
  $^2$TAPIR, Walter Burke Institute for Theoretical Physics, Mail Code
  350-17, Caltech, Pasadena, CA 91125, USA\\
  $^3$Physics Department and Astronomy Department, University of California at Berkeley, Berkeley, CA 94720, USA\\
  $^4$National Radio Astronomy Observatory, P.O. Box O, Socorro, NM 87801, USA\\
  $^5$Cahill Center for Astronomy, MC 249-17, Caltech, Pasadena, CA 91125, USA}

\begin{document}
\label{firstpage}
\maketitle

\begin{abstract}
A significant fraction of binary neutron star mergers occur in star-forming galaxies where the UV-optical and soft X-ray emission from the relativistic jet may be absorbed by dust and re-emitted at longer wavelengths. We show that, for mergers occurring in gas-rich environment ($n_{\rm H}\gtrsim 0.5\rm\, cm^{-3}$ at a few to tens of pc) and when the viewing angle is less than about $30^{\rm o}$, the emission from heated dust should be detectable by James Webb Space Telescope (JWST), with a detection rate of $\sim$$1\rm\, yr^{-1}$. The spatial separation between the dust emission and the merger site is a few to 10 milli-arcsecs (for a source distance of 150 Mpc), which may be astrometrically resolved by JWST for sufficiently high signal-noise-ratio detections. Measuring the superluminal apparent speed of the flux centroid directly gives the orbital inclination of the merger, which can be combined with gravitational wave data to measure the Hubble constant. For a line of sight within the jet opening angle, the dust echoes are much brighter and may contaminate the search for kilonova candidates from short gamma-ray bursts, such as the case of GRB 130603B.
\end{abstract}


\section{Introduction}



Observations of GW170817 provided unambiguous support that short gamma-ray bursts (GRBs) are associated with binary neutron star (bNS) mergers \citep{abbott17_GW170817, goldstein17_170817_sGRB, kasliwal17_170817_sGRB, fong19_GW170817_optical_afterglow, mooley18_superluminal_motion, margutti18_GW170817_afterglow}, in agreement with theoretical expectations \citep{eichler89_bNS_sGRBs, narayan92_bns_mergers_sGRBs, narayan01_accretion_model, Lee07_sGRB_review, nakar07_sGRB_review, berger14_sGRB_review, kumar15_GRB_review}. Most ($\gtrsim 50\%$) of the short GRBs observed so far are from late-type galaxies whereas $\sim30\%$ are from early-type galaxies \citep{fong13_sGRB_hosts}, which indicates that the probability of short GRB occurrence is not only related to the stellar mass but also strongly influenced by star formation activity \citep{berger14_sGRB_review}. Another independent evidence for short delay times between star formation and bNS merger is that $\sim30\%$ of known merging Galactic bNSs have merger time less than 100 Myr \citep[e.g.,][]{tauris15_ultra_stripped_SN, farrow19_galactic_bns}, and the cosmological bNS merger rate inferred from the Galactic bNS population \citep[e.g.,][]{kim15_galactic_bNS_rate} is comparable to that measured by LIGO-Virgo gravitational wave (GW) observations \citep{abbott20_O3a_catalog}. Furthermore, modeling of the Galactic bNS population shows that there is a statistically significant excess of systems with short delay times compared to the canonical power-law of $t^{-1}$ \citep{beniamini19_delay_time_distribution}, and such a steep delay-time distribution is also consistent with the redshift distribution of short GRBs \citep{wanderman15_sGRB_redshift_distribution}. Therefore, we expect a significant fraction of bNS mergers/short GRBs to occur in gas-rich environment, where the UV-optical and soft X-ray emission from the relativistic jet will be reprocessed by gas and dust along the beaming cone into longer wavelengths in the form of an echo. Indeed, the broadband afterglows of many short GRBs show evidence of dust extinction $A_{\rm V}\sim 0.3$--$1.5\rm\,mag$ \citep{fong15_sGRB_sample} along the jet axis, including e.g., GRB 130603B \citep{tanvir13_GRB130603B_kNe_candidate, berger13_GRB130603B_kNe_candidate, deUgartePostigo14_GRB130603B_host, fong14_GRB130603b_afterglow}.

On the other hand, it is well-known that long GRBs of core-collapse origin are located in intensively star-forming environment. A large fraction ($10$--$50\%$) of them occur in dusty environment such that their optical afterglows are attenuated by $A_{\rm V}>0.5\mr{\,mag}$ \citep{cenko09_darkGRBs, melandri12_dark_GRBs, perley13_dark_GRBs}. Thus, their afterglow emission are more susceptible to gas and dust reprocessing, which has been studied in the literature \citep{waxman00_GRB_dust_echo, esin00_GRB_dust_echo, madau00_compton_echo, fruchter01_xray_destruction, draine02_molecular_gas_absorption, perna02_dust_code, heng07_dust_echo, evans14_ulGRB_echo}. However, for the practical cases where the observer's line of sight (LOS) is close to the jet axis, the GRB-associated supernova \citep{woosley06_GRB_SN} typically outshines the dust echo in the near-infrared band.

In this paper, we propose that dust echoes from a fraction of bNS mergers, whose jet axis are generally misaligned with the observer's LOS, should be detectable by James Webb Space Telescope (JWST). This provides a new, observable electromagnetic (EM) counterpart\footnote{An earlier study on the Compton scattering echo of short GRBs found the signal to be too faint except for very nearby sources where the observer's line of sight is only slightly away from the jet beaming cone \citep{beniamini18_compton_echo}.} to bNS mergers, in addition to the kilonova/macronova \citep{li98_kilonova, kulkarni05_kilonova, metzger10_kilonova, barnes13_kilonova_high_opacity, tanaka13_kilonova_high_opacity, yu13_magnetar_merger_nova, metzger17_kilonova_review}, short GRB and synchrotron afterglow from the jet/ejecta \citep{nakar11_radio_afterglow, metzger12_EM_counterparts, piran13_EM_counterpart, hotokezaka18_ejecta_afterglow}, as well as non-thermal emission from the wind nebula of a possible long-lived NS remnant \citep{zhang13_X-ray_magnetar, metzger14_long_lived_magnetar}. We further demonstrate that, for sufficiently bright dust echoes, the separation between the echo flux centroid and the position of the merger may be astrometrically resolved by JWST, and it can be used to infer the inclination angle of the bNS orbital axis and measure the Hubble constant \citep{hotokezaka19_hubble_constant}.

\begin{figure*}
\centering
\includegraphics[width=0.6\textwidth]{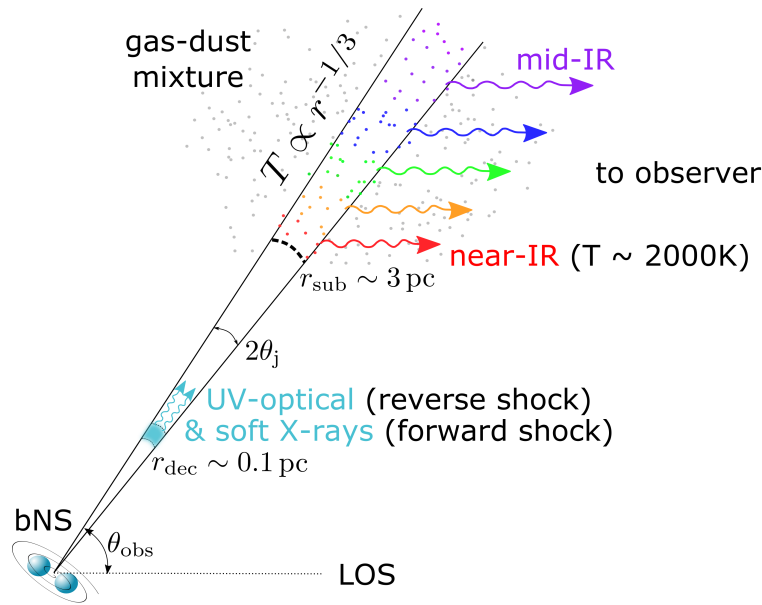}
\caption{Schematic picture of the model. A relativistic jet, launched from a binary neutron star (bNS) merger, generates bright UV-optical and soft X-ray emission when it is decelerated by the circum-stellar medium at a distance of the deceleration radius $r_{\rm dec}\sim 0.1\rm\, pc$. The jet emission is beamed in a cone of half opening angle of $\theta_{\rm j}$, and the dust grains at distances of a few to tens of parsecs within this cone are heated by the radiation to high temperatures. All dust grains are depleted below the sublimation radius $r_{\rm sub}\sim 3\rm\, pc$. The  energy absorbed by dust at $r>r_{\rm sub}$ is re-radiated in the infrared according to the local dust temperature, which roughly follows a power-law scaling with radius $T\propto r^{-1/3}$. For a line of sight (LOS) at an angle $\theta_{\rm obs}$ from the jet axis, light-travel delay causes longer wavelength dust emission to arrive at the observer at later time.
}
\label{fig:sketch}
\end{figure*}

The broad-brush picture considered in this work is shown in Fig. \ref{fig:sketch}. Before presenting the model, we list a number of potential physical sources of dust extinction in the vicinity of bNS mergers in late-type galaxies.





(1) Diffuse interstellar medium. For typical Galactic gas-to-dust ratio, the V-band extinction is related to the gas column density by $A_{\rm V}\simeq 5.3 \times 10^{-22} N_{\rm H} \rm\, mag\, cm^2\simeq 0.5\mr{\,mag}\, (\Sigma_{\rm H}/10\,M_\odot\mr{\,pc^{-2}})$, where $N_{\rm H}$ (or $\Sigma_{\rm H}$) is the number (or mass) column density, and typical non-star-burst spiral galaxies have $\Sigma_{\rm H}\sim 10\,M_\odot\mr{\,pc^{-2}}$ \citep{kennicutt12_star_formation}. The vertical scale-height of atomic gas in the disk is a few hundred parsecs, whereas the molecular gas is more vertically confined. Thus, for those short GRBs with significant V-band extinction $A_{\rm V}= 0.5\rm\, mag$ and LOS path length $\ell_{\rm H}= 300\rm\, pc$ through the gas-rich region, the mean number density of the intervening gas is $\bar{n}_{\rm H}\simeq N_{\rm H}/\ell_{\rm H}\simeq 1\mr{\,cm^{-3}}$. However, due to the inhomogeneity of the interstellar medium, the gas density in the vicinity of a bNS merger may be smaller than $\bar{n}_{\rm H}$ (see later discussions). Generally, the extinction is stronger at shorter wavelengths, with the rough scaling of $A_\lambda\propto \lambda^{-1}$ for an average Galactic LOS \citep{cardelli89_extinction_law}, so UV photons $\lambda\sim 0.1\um$ may be significantly attenuated within a distance $\ll \ell_{\rm H}$ from the source (and ionizing photons above 13.6 eV may be largely absorbed by neutral gas). UV extinction is dominated by very small grains of size $\lambda/2\pi \sim 0.02\um$. These smaller grains have lower sublimation temperature. The fact that they cool less efficiently by infrared (IR) emission makes them easier to heat up and sublimate. These two effects lead to the depletion of smaller grains in a larger volume than for bigger grains, which will later be modeled in detail.




(2) Cradle giant molecular clouds (GMCs). The average mass of molecular clouds hosting OB stars, potentially multiple generations of them, is about $3\times 10^5M_\odot$. These massive GMCs are in turn destroyed mainly by photoionization on a timescale of between 10 and 30 Myr \citep{williams97_GMC_mass_distribution, matzner02_GMC_destruction}. It is likely that some bNS mergers occur with delay times shorter than the cloud destruction time, and these mergers may still be embedded in their cradle GMC. However, the fraction of bNS mergers with delay time less than 30 Myr cannot be estimated with great confidence, given limited observational clues on the existence of such rapid mergers \citep[they may be identified by future LISA-like GW missions,][]{lau20_lisa_bNS_detections}. Theoretically, there is a ``fast channel"\footnote{Another channel is that if the second-born NS receives a large natal kick comparable to the pre-explosion binary orbital speed (a few hundred $\rm km\,s^{-1}$), there is a finite chance that the resulting bNS system may be in a highly eccentric orbit. In the high-eccentricity limit, the GW merger time distribution as a result of such a strong kick is given by $\d P/\d\, \mr{log}\,t_{\rm GW}\propto t_{\rm GW}^{2/7}$ \citep{lu21_kick_GW_merger}, and it is possible to achieve extremely short GW inspiral time $t_{\rm GW}\ll \rm\, Myr$.} where the envelope expansion of a low-mass ($2$--$4\,M_\odot$) He star leads to Roche-Lobe overflow towards a NS companion, and this leads to further shrinkage of the binary orbit before an ultra-stripped supernova and potentially very short GW inspiral time $t_{\rm GW}\lesssim\, $a few$\,\rm\, Myrs$ \citep{belczynski06_fast_channel, tauris15_ultra_stripped_SN}. The other consideration is the motion of the bNS system. Roughly 20 Myr after the binary formation (for a primary mass of $12M_\odot$ for instance), the natal kick on the first-born NS, $v_{\rm k1}$, will induce a binary center-of-mass speed of $v_{\rm k1}/10$, because the other member is still a massive main-sequence star that is 10 times more massive than the neutron star. We see that $v_{\rm k1}\lesssim 100\rm\, km\,s^{-1}$ is required for the binary to remain bound to the GMC, which has escape speed of $\sim10\rm\, km\,s^{-1}$ \citep[similar to that of the ultra-faint dwarf galaxy Rec II, which has been enriched by r-process elements likely by a bNS merger,][]{ji16_recII_enrichment}. Such a small kick is possible because the primary star has lost its hydrogen envelope due to Roche-Lobe overflow in a close binary before exploding as a Type Ib supernova, which may generate a substantially weaker NS kick than that of isolated pulsars \citep[e.g.,][]{tauris17_bNS_formation}. Then, the second-born NS causes another a center-of-mass kick of $v_{\rm k2}/2$, which could be as small as $10\rm\, km\,s^{-1}$ in the case (ultra-stripped) electron capture supernova \citep{beniamini16_bNS_kick, tauris17_bNS_formation}. Observationally, the short GRB 101219A ($z=0.718$) occurred in a host galaxy with estimated star-formation rate of $\sim 16\,M_\odot\rm\,yr^{-1}$ and stellar population age of $\sim 50\rm\,Myr$ \citep{fong13_sGRB_hosts}. The averaged star-formation rate surface density of this galaxy can  be  roughly converted to a molecular column density of $\Sigma_{\rm H_2}\sim 10^{2}\,M_\odot\rm\,pc^{-2}$ \citep{kennicutt12_star_formation}. We note that the majority of long GRBs from core-collapse of massive stars should be in this category.

(3) Unassociated GMCs. The demographics of GMCs in the Milky Way have been studied by wide field surveys of molecular emission lines. Their mass and size distributions are found to be $\d N/\d M\propto M^{-1.8}$ and $\d N/\d r_{\rm c}\propto r_{\rm c}^{-3.2}$ \citep[e.g.,][]{williams97_GMC_mass_distribution, heyer01_GMC_size_distribution}, meaning that most molecular mass [$\int \d M \, M(\d N/\d M)\propto M^{0.2}$] is in the most massive GMCs with $M_{\rm max}\sim 5\times10^6\,M_\odot$ and that most geometric cross-section for LOS-intersection [$\int \d r_{\rm c}\, r_{\rm c}^2 (\d N/\d r_{\rm c})\propto r_{\rm c}^{-0.2}$] is provided by the least massive GMCs with mass $M_{\rm min}\sim 10^3\, M_\odot$ and size $r_{\rm c,min}\sim 3\rm\, pc$. These small GMCs contains a fraction $f_{\rm M}\sim (M_{\rm min}/M_{\rm max})^{0.2}\sim 20\%$ of the total molecular mass. Suppose a bNS merger occurs in a molecular region with column density $\Sigma_{\rm H_2}=10^2\,M_\odot\,\rm pc^{-2}$ and vertical scale height of $z_{\rm H_2} = 50\rm\, pc$ \citep{heyer15_molecular_clouds_review}, then the mean number density of GMCs can be estimated by $\bar{n}_{\rm GMC}=f_{\rm M}\Sigma_{\rm H_2}/(M_{\rm min} 2z_{\rm H_2}) \sim 2\times10^{-4}\rm\,pc^{-3}$. Then, the probability that a narrow radiation beam from a GRB jet intersects a molecular cloud within a distance $r\sim 10\rm\,pc$ is given by $P\sim \bar{n}_{\rm GMC} \pi r_{\rm c,min}^2 r \simeq 6\%\, (r/10\mr{\, pc}) (\Sigma_{\rm H_2}/10^2\,M_\odot\,\rm pc^{-2})$. When such an intersection occurs, nearly all UV photons (including those above 13.6 eV) will be absorbed/scattered by dust grains, because at high gas density $n_{\rm H}\sim 10^{3}\rm\, cm^{-3}$, dust attenuation occurs within a small thickness where all the gas is ionized by a very small fraction of the ionizing photons.


The above considerations suggest that a significant fraction of bNS mergers may be surrounded by dense gas on lenthscales of a few to tens of parsecs. This motivates us to model the interaction between the jet radiation beam and the surrounding gas/dust, as well as to explore what can be learned from the detection of such dust echoes.

This paper is organized as follows. In \S \ref{sec:estimate}, we provide an order-of-magnitude estimate of the luminosity and duration of the dust echo for arbitrary viewing angles, under the assumption that the jet generates a short bright pulse of optical-UV emission within its beaming angle. Then, we show the lightcurves and flux centroid motions for a number of physically motivated cases and provide an estimate of the detection rate of dust echoes in \S \ref{sec:results}.
Detailed modeling of the dust sublimation and emission processes is presented in \S \ref{sec:model}, which is the basis of \S \ref{sec:results}.  There we focus on dust heating by UV-optical emission; the effects of X-ray photons are discussed in \S \ref{sec:X-ray}. We summarize in \S\ref{sec:summary}. The UV-optical emission from the reverse shock formed when the jet runs into the circum-stellar medium is calculated Appendix \ref{sec:reverse_shock}. We use the convention $Q = 10^x Q_x$ in CGS units.

\section{Order-of-magnitude Estimate}\label{sec:estimate}

In this section, we estimate the luminosity and duration of the dust echo in a number of simple scenarios.

The first case is that the observer's viewing angle $\theta_{\rm obs}$ wrt. the jet axis is much smaller than the opening angle of the jet $\theta_{\rm j}$ --- the LOS is effectively coincident with the jet axis. As a result of light travel delay, a spherical shell of dust at radius $r$ heated from an isotropic short burst of radiation produces a square-function echo light curve of duration $2r/c$ (this is called the ``response function''). Here, the jet emission only illuminates a small patch of solid angle $\pi \theta_{\rm j}^2$ on the sphere, so the duration of the echo is $r\theta_{\rm j}^2/(2c)$ but the luminosity is the same as in the case of a full sphere. Then we consider a large number of spherical shells, each of which has absorption optical depth $\d\tau_{\rm abs}$ (the fraction of the incident UV-optical energy $E_{\rm UV}$ absorbed by this shell), and the total bolometric echo luminosity contributed by dust grains in all shells is given by
\begin{equation}
    L_{\rm d}^{\rm on\mbox{-}axis} \sim \int_{r_{\rm sub}} \d \tau_{\rm abs} {E_{\rm UV} \mr{e}^{-\tau_{\rm ext}} \over 2r/c},
\end{equation}
where $\tau_{\rm ext}\simeq 2\tau_{\rm abs}$ is the extinction (absorption plus scattering) optical depth below radius $r$, and $r_{\rm sub}$ is the sublimation radius below which all dust grains are heated to above the sublimation temperature $T_{\rm sub}\sim 2200\rm\, K$ (see \S \ref{sec:model}) and have evaporated. Here for simplicity, we ignore the attenuation due to gas ionization, because the isotropic equivalent energy it takes to ionize all the gas up to radius $r$ is $\sim10^{47}\mr{\,erg}\, (n_{\rm H}/1\mr{\,cm^{-3}}) (r/3\,\mr{pc})^3\ll E_{\rm UV}\sim 10^{50}\rm\, erg$. As shown in Appendix \ref{sec:reverse_shock}, when the jet is decelerated by the gas in the circum-stellar medium (CSM) at a distance $\sim 0.1\rm\, pc$ (the deceleration radius), the electrons accelerated by the reverse shock generate bright synchrotron emission in the UV-optical frequency range with isotropic equivalent energy $E_{\rm UV}\sim 10^{50}\rm\, erg$, for an isotropic equivalent jet kinetic energy $E_{\rm j}\sim\,$a few$\times 10^{52}\rm\, erg$ \citep[as inferred from the luminosity function of short GRBs, e.g.,][]{beniamini19_sGRB_170817} and CSM density $\gtrsim 0.1\rm\, cm^{-3}$. For isotropic equivalent UV luminosity $L_{\rm UV} = E_{\rm UV}/t_{\rm UV}$ and typical grain size $a = 0.1\,\um$, the sublimation radius is estimated to be $r_{\rm sub}\simeq 3\mr{\,pc}\, L_{\rm UV, 48}^{1/2} a_{\rm 0.1\um}^{-1/2}$ (see \S \ref{sec:model}). At much larger distances $r\gg r_{\rm sub}$, the grain temperature drops as $T\propto r^{-1/3}$ (see \S \ref{sec:model}) and the peak frequency of the dust emission goes to longer and longer wavelengths.

In the optically thin limit $\tau_{\rm ext}\ll 1$, even though each radial shell with equal intervals in $\mr{log}(r)$ gives the same amount of bolometric echo luminosity, the contribution to the near-IR ($\lambda_{\rm obs}\sim\,$a few $\um$) flux is dominated by the region near the sublimation radius. Thus, the specific luminosity of the dust echo near the peak wavelength $\lambda_{\rm obs}\sim h c/3kT_{\rm sub}\sim 2\,\um (T_{\rm sub}/2200\mr{\,K})^{-1}$ is given by
\begin{equation}\label{eq:on_axis_luminosity}
\begin{split}
    L_{\rm d,\nu}^{\rm on\mbox{-}axis}&\sim {h \over 3kT_{\rm sub}} \min\left[\tau_{\rm abs}({r_{\rm sub}}), {1\over 2}\right] {E_{\rm UV}\over 2r_{\rm sub}/c}\\
    &\sim (6\times10^{26}\mr{\, erg\,s^{-1}\,Hz^{-1}})\, \min\left[2\tau_{\rm abs}({r_{\rm sub}}), 1\right] L_{\rm UV,48}^{1/2} t_{\rm UV,2}\, a_{0.1\um}^{1/2},
\end{split}
\end{equation}
where we have used fiducial UV-optical luminosity $L_{\rm UV} = 10^{48}\rm\, erg\,s^{-1}$ and duration $t_{\rm UV}=100\rm\, sec$ (cf. Appendix \ref{sec:reverse_shock}). Note that eq. (\ref{eq:on_axis_luminosity}) applies to both optically thin ($\tau_{\rm abs}\ll 1$) and optically thick ($\tau_{\rm abs}\gg 1$) dust columns. The duration of the dust echo is given by
\begin{equation}\label{eq:onaxis_peak_time}
    t_{\rm d, obs}^{\rm on\mbox{-}axis}\sim {\theta_{\rm j}^2\over 2} {r_{\rm sub}\over c} \sim (9\mr{\,d})\, (\theta_{\rm j}/4^{\rm\,o})^2L_{\rm UV, 48}^{1/2} a_{\rm 0.1\um}^{-1/2}.
\end{equation}
The above estimate shows that dust echo may be responsible for the famous kilonova candidate from GRB 130603B \citep{tanvir13_GRB130603B_kNe_candidate, berger13_GRB130603B_kNe_candidate}, which was based on one H-band (rest-frame $\lambda_{\rm obs}=1.2\um$) detection with $L_\nu \simeq 8\times10^{26}\rm\, erg\,s^{-1}\,Hz^{-1}$ at $t_{\rm obs} = 6.9\rm\ d$ in the host-galaxy rest frame (redshift $z=0.356$). This will be further discussed in \S \ref{sec:offaxis}.

The second case is that the observer's viewing angle is much larger than the jet opening angle, $\theta_{\rm obs}\gg \theta_{\rm j}$. In this case, the dust echo arrives at the observer after a delay
\begin{equation}\label{eq:offaxis_peak_time}
    t_{\rm d, obs}^{\rm off\mbox{-}axis} \sim {r_{\rm sub} \over c}(1-\cos\theta_{\rm obs})\simeq 1.3\mr{\,yr}\, (\theta_{\rm obs}/30^{\rm o})^2 L_{\rm UV, 48}^{1/2} a_{\rm 0.1\um}^{-1/2},
\end{equation}
where the second expression is valid for $\theta_{\rm obs}\lesssim 1\rm\, rad$ (assumed to be true hereafter). The peak flux in the off-axis case is much smaller than the on-axis case by a factor that is of the order $(\theta_{\rm j}/\theta_{\rm obs})^2$, for the following two reasons. First, only a fraction of the entire azimuth around the LOS is occupied by the jet-illuminated patch, $\theta_{\rm j}/[2\pi \sin(\theta_{\rm obs})] \simeq 0.02\, (\theta_{\rm j}/4^{\rm o}) (\theta_{\rm obs}/30^{\rm o})^{-1}$. Second, since the echo from an infinitesimal radial shell at radius $r_{\rm sub}$ lasts for a duration of $\Delta t_{\rm d, obs} \simeq (r_{\rm sub}/c) \theta_{\rm j} \sin \theta_{\rm obs}$, the total flux only comes from the contribution from a smaller radial thickness of $\Delta r/r_{\rm sub} \simeq \theta_{\rm j} \sin \theta_{\rm obs}/(1-\cos\theta_{\rm obs}) \simeq 1/4\, (\theta_{\rm j}/4^{\rm o}) (\theta_{\rm obs}/30^{\rm o})^{-1}$. Combining this two factors, we obtain the peak flux for the off-axis case
\begin{equation}
\begin{split}
    L_{\rm d,\nu}^{\rm off\mbox{-}axis} &\sim {\theta_{\rm j} \over 2\pi \sin\theta_{\rm obs}} {\theta_{\rm j} \sin \theta_{\rm obs} \over 1-\cos\theta_{\rm obs}} L_{\rm d,\nu}^{\rm on\mbox{-}axis} \simeq {h\over 3kT_{\rm sub}} \mr{min}\left[\tau(r_{\rm sub}), {1\over 2}\right] {E_{\rm UV}\theta_{\rm j}^2 \over 4\pi (1-\cos\theta_{\rm obs}) r_{\rm sub}/c}\\
    & \sim (3\times10^{24}\mr{\, erg\,s^{-1}\,Hz^{-1}})\, (\theta_{\rm j}/4^{\rm o})^2 (\theta_{\rm obs}/30^{\rm o})^{-2} \min\left[2\tau_{\rm abs}({r_{\rm sub}}), 1\right] L_{\rm UV,48}^{1/2} t_{\rm UV,2}\, a_{0.1\um}^{1/2},
\end{split}
\end{equation}
Another way of understanding the $\theta_{\rm obs}^{-2}$ scaling of the peak luminosity is that the total energy from dust emission is independent of the viewing angle and that the flux is spread out over a longer time for larger viewing angles with the dependence $L_{\rm d,\nu}^{\rm off\mbox{-}axis}\propto (1-\cos\theta_{\rm obs})^{-1}\approx \theta_{\rm obs}^{-2}/2$. After reaching a peak at $t_{\rm d,obs}^{\rm off\mbox{-}axis}$ (eq. \ref{eq:offaxis_peak_time}), the flux will stay near the peak for a duration comparable to the peak time, and then the flux will decline gradually, as the contribution from colder dust at larger radii $r\gg r_{\rm sub}$ becomes dominant. Generally, at longer wavelengths, the flux reaches the peak at later times and the post-peak decline is slower.

So far we have thrown away the dust-scattered light, which may be observable in the optical band by ground-based observatories (e.g., future Extremely Large Telescopes). Since the scattering cross-sections are similar to absorption, we expect the specific luminosity or photon number flux of the scattered light in the optical band (at wavelength $\lambda_{\rm opt}$) to be smaller than that from dust emission by a factor of $(\lambda_{\rm opt}/\lambda_{\rm IR}) (E_{\rm opt}/E_{\rm UV}) \sim 0.1$, where we have taken $\lambda_{\rm opt}\sim 0.6\,\um$ (V-band), $\lambda_{\rm IR}\sim 2\,\um$ (peak wavelength for dust near the sublimation temperature), and a fraction $(E_{\rm opt}/E_{\rm UV})\sim 1/3$ of the jet emission in the $0.02$--$1\um$ range to be in the optical band near $\lambda_{\rm opt}$. At small viewing angles $\theta_{\rm obs}\lesssim 30^{\rm o}$, this flux-density reduction is compensated (by a factor of $\sim 2$) by the fact that forward-scattering is slightly favored \citep{draine11_book}. Additionally, the kilonova emission \citep[especially in the case of a long-live NS remnant][]{yu13_magnetar_merger_nova, metzger14_long_lived_magnetar} may also contribute to the flux of the scattering echo, whose spectrum may have line features from atomic transitions. However, there may be additional dust extinction along the LOS which further reduces the flux of the scattered light. In this work, we focus on the infrared emission from the dust heated by UV-optical emission from the jet and do not consider scattered light.

Finally, the flux centroid of the dust emission is located at a projected distance of $r_{\rm sub}\sin\theta_{\rm obs}$ from the center of explosion. For a source (angular-diameter) distance of $D$, this gives an angular separation of
\begin{equation}
    \tilde{\theta}_{\rm proj} \sim {r_{\rm sub}\sin\theta_{\rm obs} \over D} \simeq 3\mr{\,mas}\, (\theta_{\rm obs}/30^{\rm o}) L_{\rm UV, 48}^{1/2} a_{\rm 0.1\um}^{-1/2} (D/100\rm\,Mpc)^{-1}.
\end{equation}
In our numerical model, we find the projected angular separation to be slightly larger than the estimate above, because of the flux contribution from radii slightly larger than $r_{\rm sub}$. The spatial extent of the dust echo is given by the transverse size of the illuminated patch $\theta_{\rm j} r_{\rm sub}$, which gives an angular size of less than 1 mas. This means that the dust echo will be almost like a point source. Since the physical position of the dust emission region moves at speed of light, the observed flux centroid of the dust echo will have apparent proper motion (in units of speed of light)
\begin{equation}\label{eq:bapp}
    \beta_{\rm app} = {\sin\theta_{\rm obs} \over 1-\cos\theta_{\rm obs}} \simeq 3.7\, (\theta_{\rm obs}/30^{\rm o})^{-1},
\end{equation}
which should in principle be measurable given two epochs of astrometric dust echo data, or one epoch of dust echo data plus an earlier epoch of kilonova emission (as the reference point). 



\section{Dust Echo Lightcurves, Flux Centroid Positions, and Detection Rate}\label{sec:results}
We defer the detailed description of our numerical model to \S \ref{sec:model}. In this section, we show the dust echo lightcurves and flux centroid positions for a number of representative cases (the data can be downloaded from this URL\footnote{\href{https://github.com/wenbinlu/dustecho.git}{https://github.com/wenbinlu/dustecho.git}}). Then, we provide an estimate of the detection rate of dust echoes by JWST.

We consider a physical situation where the UV-optical emission from the jet is a square pulse of isotropic equivalent luminosity $L_{\rm UV} = 3\times 10^{47}\rm\, erg\, s^{-1}$ and duration $t_{\rm UV} = 300\rm\,s$. The spectrum is taken to be a power-law from synchrotron emission with $L_\nu\propto \nu^{(1-p)/2}$ and $p=2.2$, and the normalization is $L_{\nu_{\rm max}}= (3-p) L_{\rm UV}/(2  \nu_{\rm max})$ at maximum frequency $h\nu_{\rm max}=50\rm\, eV$. As shown in Appendix \ref{sec:reverse_shock}, these parameters are expected from a bNS merger like GW170817 with isotropic equivalent jet energy of $E_{\rm j}\sim\,$a few$\times 10^{52}\rm\, erg$ \citep[e.g.,][]{lazzati18_GW170817_jet, mooley18_superluminal_motion, gill18_170817_jet, beniamini19_sGRB_170817, hajela19_170817_fits} but embedded in a denser CSM of density $\gtrsim 10^{-1}\rm\, cm^{-3}$ in a late-type galaxy \citep[see][for constraints on CSM densities for a sample of short GRBs]{oconnor20_CSM_density_constraint}. Note that the jet optical emission depends on the CSM density at a distance of $\sim 0.1\rm\,pc$, which could be unrelated to the density of the gas at a few to tens of parsecs that is responsible for the dust echo. Photons above $50\rm\, eV$ may contribute comparable amount of heating (despite reduced absorption cross-sections, see Fig. \ref{fig:extinction}), but we do not consider soft X-ray photons in this work because rapid cooling of the radiating electrons accelerated by the reverse shock may suppress the emission above 50 eV. The effects of X-ray photons will be discussed in \S \ref{sec:X-ray}. We assume the half opening angle of the jet emission to be $\theta_{\rm j} = 4^{\rm o}$, which is consistent with the jet in GW170817 \citep{mooley18_superluminal_motion, mooley18_jet_opening_angle}. 

We calculate the sublimation radius for dust grains of different sizes and the extinction in an iterative way, and then the volumetric emissivity is calculated from the dust temperatures for all grain sizes, and finally, we account for light-travel delays and obtain the lightcurve of dust echo from a given observer's viewing angle $\theta_{\rm obs}$ and wavelength $\lambda_{\rm obs}$. Below, we show the results for the off-axis ($\theta_{\rm obs}\gg \theta_{\rm j}$) case in \S \ref{sec:offaxis} and then the on-axis ($\theta_{\rm obs}< \theta_{\rm j}$) case in \S \ref{sec:onaxis}.

\subsection{Off-axis Case}\label{sec:offaxis}
\begin{figure*}
\centering
\includegraphics[width=0.9\textwidth]{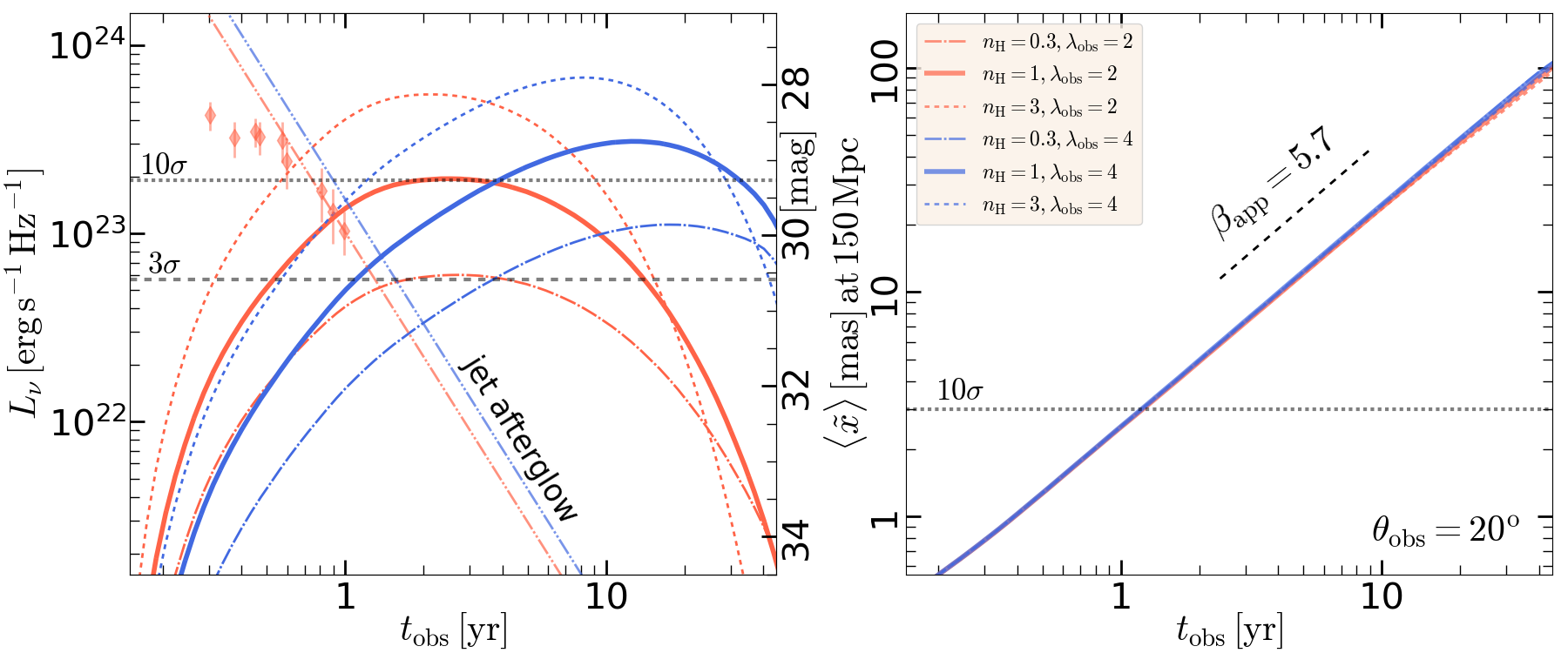}
\caption{\textit{Left panel}: Lightcurves of the dust echo emission at different wavelengths and for three different scenarios of hydrogen number densities $n_{\rm H} = 0.3\rm\, cm^{-3}$ (dash-dotted lines), $1\rm\, cm^{-3}$ (solid lines), and $3\rm\, cm^{-3}$ (dashed lines). The grey dotted (dashed) line denotes $10\sigma$ ($3\sigma$) sensitivity of JWST for 20 ksec exposure time at a luminosity distance of $150\rm\, Mpc$. Different line colors corresponding to wavelengths of $\lambda_{\rm obs}=2\, \um$ (red) and $4\,\um$ (blue). For all cases, we have fixed the observer's viewing angle $\theta_{\rm obs}=20^{\rm o}$, jet opening angle $\theta_{\rm j}=4^{\rm o}$, and dust-to-gas ratio parameter $n_0/n_{\rm H} = 1.45\times10^{-15}$ (for a typical LOS in the Milky Way). The emission from the jet is taken to be a square pulse with isotropic equivalent UV-optical luminosity $L_{\rm UV} = 3\times 10^{47}\rm\, erg\, s^{-1}$, spectrum $L_\nu \propto \nu^{(1-p)/2}$ ($p=2.2$), and duration $t_{\rm UV}=300\rm\, s$. The red diamonds show the afterglow lightcurve of GW170817, which are obtained by scaling the R-band flux measurements to wavelength of 2$\um$ according to the known broad-band power-law spectrum \citep{fong19_GW170817_optical_afterglow}. The red and blue dashed-double-dotted lines denote $F_\nu\propto t_{\rm obs}^{-p}$ extrapolations of the GW170817 afterglow at $\lambda_{\rm obs}=2$ and $4\um$, respectively. For any CSM density, the late-time jet afterglow emission from a GW170817-like bNS merger should be near or to the left of these two lines. \textit{Right panel}: Time evolution of the angular separation between the dust echo flux centroid and the center of explosion, for an assumed angular diameter distance of $150\rm\, Mpc$. The results in all cases agree with the analytical result of apparent speed $\beta_{\rm app}=\sin\theta_{\rm obs}/(1-\cos\theta_{\rm obs})$. The astrometric precision of JWST is estimated to be about $3\rm\,mas$ for $\rm SNR=10$, which is shown by a horizontal grey dotted line.
}
\label{fig:echo_theobs20}
\end{figure*}

\begin{figure*}
\centering
\includegraphics[width=0.9\textwidth]{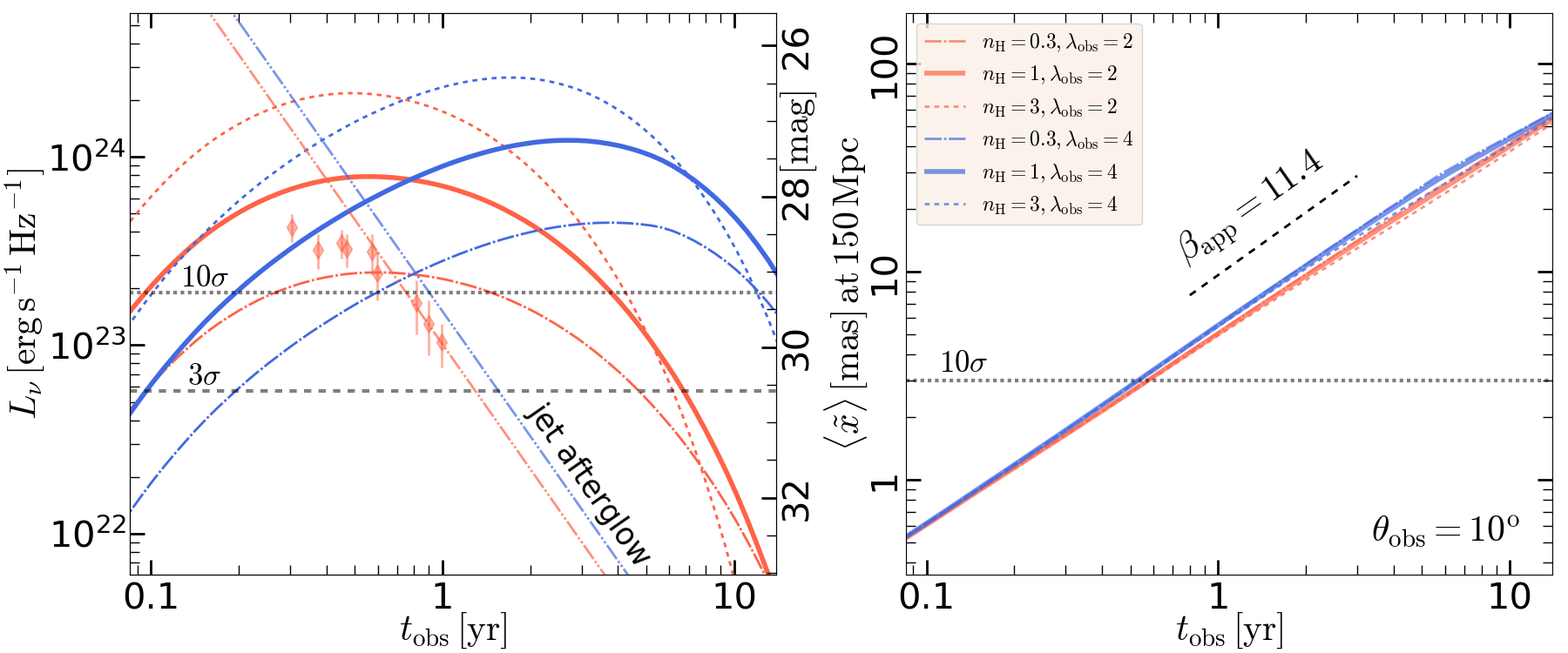}
\caption{Same as Fig. \ref{fig:echo_theobs20} but for a viewing angle of $\theta_{\rm obs}=10^{\rm o}$ and the LOS is not far from the jet beaming cone of opening angle $\theta_{\rm j}=4^{\rm o}$.
The flux reaches the peak at an earlier time than the $\theta_{\rm obs}=20^{\rm o}$ case and the peak flux is higher.
}
\label{fig:echo_theobs10}
\end{figure*}

\begin{figure*}
\centering
\includegraphics[width=0.9\textwidth]{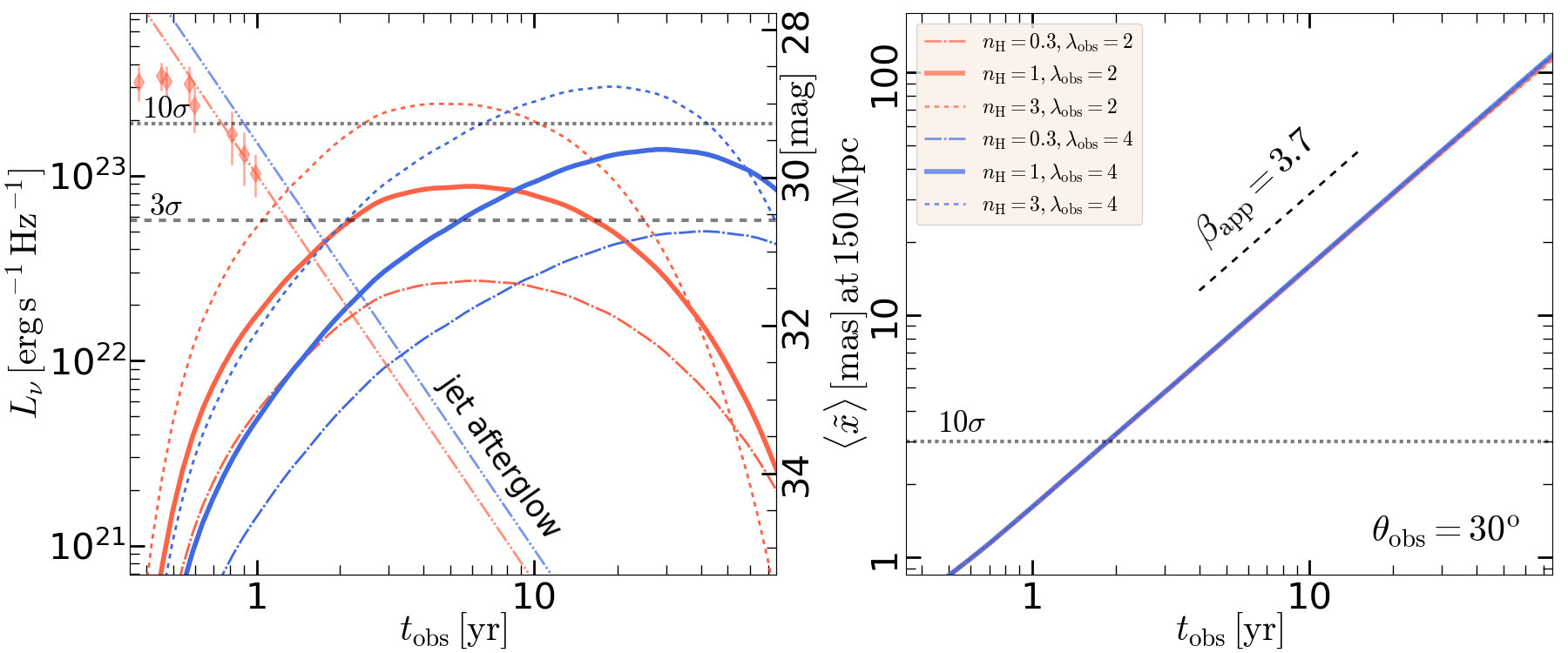}
\caption{Same as Fig. \ref{fig:echo_theobs20} but for a viewing angle of $\theta_{\rm obs}=30^{\rm o}$. The flux reaches the peak at a later time than the $\theta_{\rm obs}=20^{\rm o}$ case and the peak flux is lower.
}
\label{fig:echo_theobs30}
\end{figure*}

The results of the off-axis case are shown in Figs. \ref{fig:echo_theobs20}, \ref{fig:echo_theobs10}, and \ref{fig:echo_theobs30} for three different viewing angles of $\theta_{\rm obs}=20^{\rm o}$, $10^{\rm o}$ and $30^{\rm o}$ respectively, and in each figure, we consider three different hydrogen number densities $n_{\rm H} = 0.3,$ 1, 3$\,\rm cm^{-3}$ for the gas-dust mixture and different observing wavelengths $\lambda_{\rm obs}=2,$ $4\,\rm \um$. We find that the dust echo brightens on a timescale of months to a few years after the merger, strongly depending on the viewing angle. For a source distance of 150 Mpc, gas density $n_{\rm H} \gtrsim  0.5\rm\, cm^{-3}$ and viewing angle $\lesssim 30^{\rm o}$, the peak flux is photometrically detectable at $3\sigma$ level by JWST with $20\rm\,ksec$ exposure.

We also show that the late-time jet afterglow emission is subdominant compared to the dust echo, regardless of the density of the CSM that the jet runs into. This is because, at a higher CSM density, the jet decelerates more rapidly, the afterglow flux peaks earlier and at a larger value, but then the flux drops rapidly as $F_{\nu}\propto t_{\rm obs}^{-p}$ after the peak time, where $p$ is the power-law index of the electron Lorentz factor distribution. Let us consider a jet with a fixed total energy interacting with CSM of different densities. For an off-axis LOS $\theta_{\rm obs}\gg \theta_{\rm j}$, the time of the afterglow flux peak has the analytical scaling $t_{\rm obs,p}\propto n^{-1/3}\theta_{\rm obs}^{2}$, the peak flux scales as $F_{\nu,\rm p}\propto n^{(p+1)/4}\theta_{\rm obs}^{-2p}$ \citep{nakar02_orphan_afterglow, gottlieb19_offaxis_afterglow}. The afterglow flux after the peak is given by $F_\nu\simeq F_{\nu,\rm p} (t_{\rm obs}/t_{\rm obs,p})^{-p}\propto n^{(3-p)/12} t_{\rm obs}^{-p}$, which depends on the CSM density $n$ extremely weakly (for $p=2.2$, the dependence is $n^{1/15}$) and is independent of the viewing angle $\theta_{\rm obs}$. Therefore, for a jet with similar energy and structure as GW170817 interacting with higher-density CSM, we expect the afterglow lightcurve after the flux peak to be close to the extrapolation of the post-peak segment of the GW170817 afterglow. Such an extrapolation is shown as thin dash-double-dotted lines in Figs. \ref{fig:echo_theobs20}, \ref{fig:echo_theobs10}, and \ref{fig:echo_theobs30}, where the flux normalization at two different wavelengths $\lambda_{\rm obs}=2\um$ (red) and $4\um$ (blue) are obtained from the precisely measured power-law spectrum of GW170817 afterglow $F_\nu\propto \nu^{-0.58}$ \citep[e.g.,][]{margutti18_GW170817_afterglow, troja2020, makhathini20_full_afterglow}. This means that, for any CSM density, the late-time jet afterglow emission should be near or to the left of the extrapolated lines.

\begin{figure*}
\centering
\includegraphics[width=0.5\textwidth]{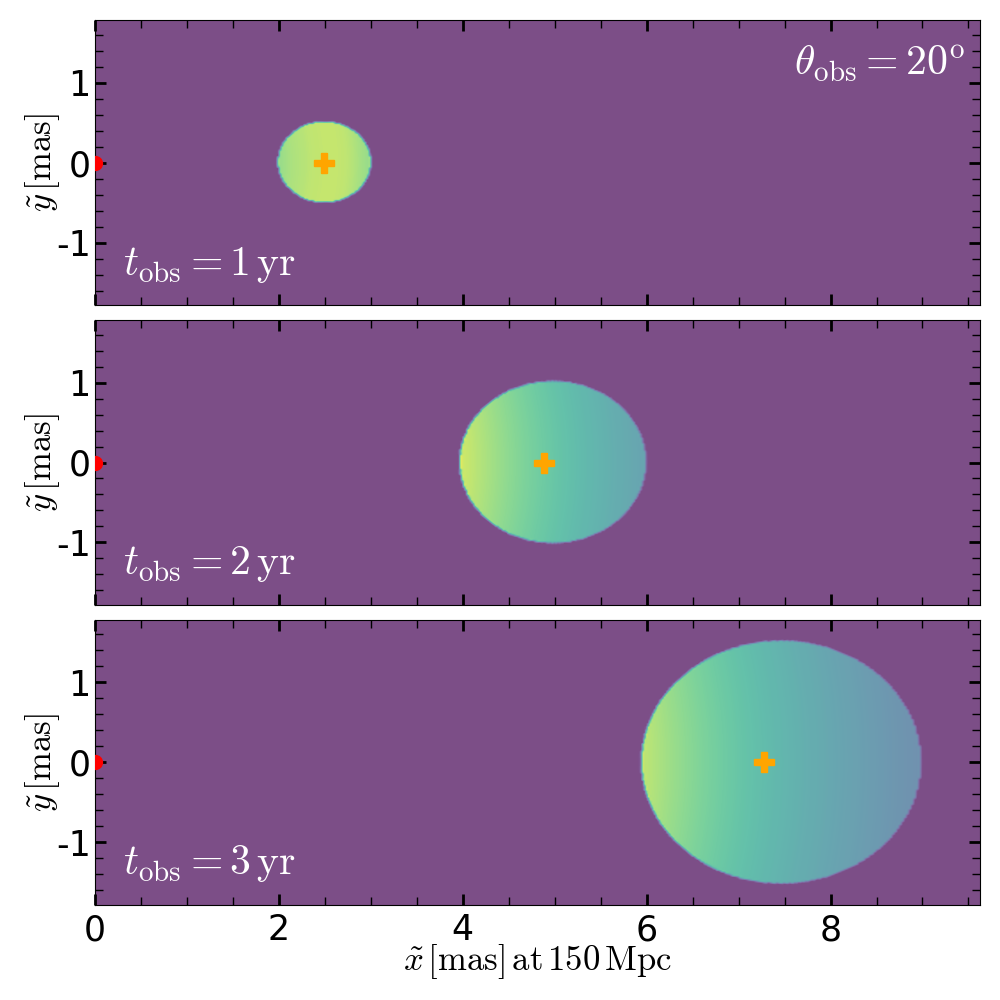}
\caption{Multi-epoch intensity maps of the dust echo, showing the flux centroid (marked by an orange cross in each panel) moving across the sky at a superluminal speed. For this case, we consider hydrogen density $n_{\rm H}=1\rm\, cm^{-3}$ (unimportant, as long as the dust column is optically thin), viewing angle $\theta_{\rm obs}=20^{\rm o}$, observing wavelength $\lambda_{\rm obs}=2\um$, and source angular diameter distance of $150\rm\, Mpc$. The position of the center of explosion (marked by a red circle) is at $(\tilde{x}=0, \tilde{y}=0)$, which can be determined to a precision much better than $1\rm\, mas$, provided that the bright kilonova emission is detected by JWST within the first month after the merger. The astrometric precision of JWST is estimated to be about $3\rm\,mas\,(SNR/10)^{-1}$, so the proper motion of the dust echo may be resolved at $t_{\rm obs}\gtrsim 2\rm\, yr$ after the merger.
}
\label{fig:intensity}
\end{figure*}

On the right panels of the figures, we show the time evolution of the position of the flux centroid projected on the sky. For a source distance of 150 Mpc, the angular separation between the echo flux centroid and the center of explosion is a few to ten milli-arcseconds, which may be resolvable by JWST when the signal-noise-ratio (SNR) is above 10. The time-dependent intensity maps of the dust emission for a representative case are shown in Fig. \ref{fig:intensity}. The astrometric error of JWST is estimated to be $1/\rm SNR$ times the half-width-half-maximum (HWHM) of the point spread function and the HWHM near $2\um$ is about $30\rm\, mas$ (which is also the pixel scale of NIRCam), so the $1\sigma$ astrometric precision is given by $3\rm\,mas\, (SNR/10)^{-1}$ (Mooley et al. in prep). The echo has superluminal apparent motion at a speed given by eq. (\ref{eq:bapp}), which is independent of the dust grain model and the details of the jet optical emission. Thus, measuring this apparent speed provides a model-independent way of obtaining the LOS inclination angle, which can be used to measure the Hubble constant \citep{hotokezaka19_hubble_constant}. 

It should also be noted that, since we are mainly interested in the cases with $\theta_{\rm obs}\lesssim 30^{\rm o}$, the dust echo from the counter jet (which propagates along the jet axis away from the observer) is expected to be temporally separate from that from the forward jet (propagating towards the observer) by $2r_{\rm sub}/c\sim 20\rm\, yr$ or longer. Furthermore, since the dust echo from the counter jet is spread out over a much longer duration of $\sim r_{\rm sub}/c$, the flux is much smaller than the forward jet echo by at least an order of magnitude \citep[e.g.,][]{heng07_dust_echo}. Therefore, astrometric measurement of the flux centroid motion within 20 years after the merger is practically unaffected by the counter jet. 

In the following, we provide a rough estimate of the rate of GW-selected bNS mergers that have detectable dust echoes. The two main factors affecting the detectability of dust echoes are the gas density near the jet axis $n_{\rm H}$ at distances of a few to tens of parsecs and the viewing angle $\theta_{\rm obs}$.

The interstellar medium is known to be highly inhomogeneous, with $\sim$half the volume occupied by very tenuous hot ionized medium \citep[$T_{\rm HIM}\sim 10^6\rm\, K$,][]{mckee77_multiphase_ISM} and the other $\sim$half is the warm neutral medium \citep[WNM, $T_{\rm WNM}\simeq 8000\rm\, K$,][]{wolfire03_WNM}. It is likely that a fraction of bNS mergers, those with very short delay times $\lesssim 30\rm\, Myr$, may be preferentially located in denser regions (the cold neutral medium), but here we conservatively assume that bNS mergers are uniformly distributed within the gaseous disk. For the $\sim$half of them embedded in WNM, the density of the surrounding gas can be estimated following the model by \citet{ostriker10_equilibrium_model} based on thermal equilibrium as well as dynamical equilibrium in the direction perpendicular to the galactic disk. If the gravity in the disk is dominated by stars and dark matter, the thermal pressure of WNM is given by \citep{ostriker10_equilibrium_model}
\begin{equation}
    P_{\rm th} \simeq \Sigma_{\rm H} \left(2\pi G \zeta_{\rm d} c_{\rm w}^2 f_{\rm w} \rho_{\rm sd} \over \alpha\right)^{1/2},
\end{equation}
where $\Sigma_{\rm H}$ is the column density of diffuse HI gas, $\zeta_{\rm d} \simeq 0.33$ is a constant that depends weakly on the vertical distribution of the gas, $c_{\rm w}\simeq 8\rm\, km\, s^{-1}$ is the thermal sound speed (since the $T_{\rm WNM}$ is set by thermal balance and is insensitive to local galactic conditions), $f_{\rm w}\sim 0.5$ is the fraction of HI in WNM phase, $\rho_{\rm sd}$ is the density of stars and dark matter in the disk mid-plane, and $\alpha\sim 5$ is the ratio of total pressure (turbulent plus magnetic) to thermal pressure. Adopting the relation between V-band extinction and gas column density $A_{\rm V}\simeq 0.5\mr{\,mag} (\Sigma_{\rm H}/10\, M_{\odot}\, \rm pc^{-2})$ and $\rho_{\rm sd}\sim 0.1\, M_\odot\, \rm pc^{-3}$ (as appropriate for the Galactic inner disk\footnote{\citet{mckee15_density_solar_neighborhood} provided an estimate of $\rho_{\rm sd}\simeq 0.05\,M_\odot\,\rm pc^{-3}$ in the solar neighborhood. Studies of Galactic supernova remnants show that most supernovae occur near galactocentric radius of $\sim$5 kpc \citep{green16_SNR_distribution}, where the density of stars and dark matter is higher than that of the solar neighborhood by a factor of $\sim$2.}), we obtain the typical density of the WNM
\begin{equation}
    n_{\rm WNM}\simeq 0.8\, \mr{cm^{-3}} {A_{\rm V}\over \mr{mag}} \left(\rho_{\rm sd} \over 0.1\,M_\odot\rm\, pc^{-3}\right)^{1/2}.
\end{equation}
We also note that, if the self-gravity of the gas in the disk is significant, then the density of the WNM will be larger.

Therefore, we see that bNS mergers with extinction\footnote{Note that, technically speaking, $A_{\rm V}$ is the extinction for a line of sight that goes through the entire disk in the vertical direction. Suppose a bNS merger occurs, on average, near the disk mid-plane, then the line of sight only goes through half of the disk, but inclinations away from the vertical direction makes the path length $\sim$twice longer than the half thickness.} $A_{\rm V}\gtrsim 0.5\rm\, mag$, LOS viewing angle $\theta_{\rm obs}\lesssim 30^{\rm o}$, and distance $D\lesssim 150\rm\, Mpc$ should yield dust echoes detectable by JWST. These requirements are described by: (1) $P_1\sim 20$--$50\%$ of bNS mergers occur in star-forming galaxies with significant dust extinction ($A_{\rm V}\gtrsim 0.5\rm\, mag$) along the jet axis \citep{fong13_sGRB_hosts}, (2) $P_2\sim 0.5$ of the gaseous disk volume is occupied by WNM,  (3) $P_3\simeq 40\%$ of GW-selected mergers have viewing angle less than 30$^{\rm o}$ \citep{schutz11_inclination_distribution}. Thus, we infer the detection rate of dust echoes within 150 Mpc to be in the range $0.5$--$1.4\rm\, yr^{-1} (\mc{R}_{\rm bNS}/10^{3}\rm\, Gpc^{-3}\,yr^{-1})$, where $\mc{R}_{\rm bNS}$ is the (highly uncertain) bNS merger rate\footnote{We note that \citet{Abbott20_bNS_rate} inferred $\mc{R}_{\rm bNS}\sim 300\rm\, Gpc^{-3}\,yr^{-1}$ based on a uniform prior of $[1, 2.5]M_\odot$ on the masses of the merging NSs. If most NS masses are closer to $1.35M_\odot$ \citep[as inferred from the Galactic bNS population][]{farrow19_galactic_bns} which correspond to a much smaller detectable volume than the $2.5M_\odot$ ones, then the rate from \citet{Abbott20_bNS_rate} is likely an underestimate. The bNS merger rate will be accurately measured by the upcoming O4 run \citep[see][]{abbott18_LIGO_horizon}.}.

There is also an interplay between viewing angle and source distance. Since the peak flux of dust echo scales as $F_{\rm peak}\propto \theta_{\rm obs}^{-2}\propto D^{-2}$, we see that the detection horizon for dust echoes scales as $D_{\rm max}\propto \theta_{\rm obs}^{-1}$. For small viewing angles $\theta_{\rm obs}\lesssim 30^{\rm o}$, the detection rate scales as $D_{\rm max}^3\theta_{\rm obs}^{1.7}\propto D_{\rm max}^{1.3}$, where the factor of $\theta_{\rm obs}^{1.7}$ comes from GW selection \citep{schutz11_inclination_distribution}. Thus, events at larger distances up to the GW detection horizon of $\sim$$400\rm\, Mpc$ for the upcoming O4 observing run \citep{chen21_detection_horizon} increases the detection rate of dust echoes by a factor of about 3.6 --- the total detection rate of dust echoes by JWST is in the range $1.7$--$5\rm\, yr^{-1} (\mc{R}_{\rm bNS}/10^{3}\rm\, Gpc^{-3}\,yr^{-1})$. Furthermore, the kilonova emission (especially the blue component), as well as the jet afterglow, from events with smaller inclination angles are generally brighter \citep{kawaguchi20_kN_viewing_angle, darbha20_kN_viewing_angle, zhu20_kN_viewing_angle, gottlieb19_offaxis_afterglow}. Thus, the inclinations of the events with precise localization by their EM counterparts are biased towards lower angles, and hence the fraction of them with detectable dust echoes is higher than the purely GW-selected sample. However, we note that, for source distances larger than 150 Mpc, it may be very challenging to measure the angular separation between the merger position and the flux centroid of the dust echo.

We conclude that it should be possible for JWST to detect the dust echoes from bNS mergers and that the detection rate is of the order once per year. We suggest the following strategy: (1) after the GW alert of each bNS merger, conduct rapid ground-based wide-field optical and/or infrared surveys \citep[e.g.,][]{andreoni20_optical_limits_to_GW} to identify the host galaxy and focus on the cases of star-forming galaxies (non-star-forming-host cases may be interesting for other purposes); (2) use JWST to take a deep exposure of the kilonova emission within the first month of the merger \citep{kasliwal19_170817_spitzer}, with an exposure time such that photometric SNR is high enough ($\gg 10$) to accurately establish the astrometric position of the merger, which is registered in the GAIA frame; (3) when the jet afterglow has faded away, take one more deep ($\sim$20 ksec) JWST exposure to search for the dust echo; (4) in the case of a successful detection of the dust echo, take additional exposures to increase the SNR to greater than 10 and then look for astrometric shift between the echo and the position of the kilonova.

\subsection{On-axis Case}\label{sec:onaxis}
When the observer's LOS is close to the jet axis (in the case where a short GRB is observed), the dust echoes may be sufficiently bright in the near-IR to be confused with kilonovae. Here we show that, the H-band excess from GRB 130603B at host-galaxy rest-frame time $t_{\rm obs}=6.9\rm\,d$ \citep{berger13_GRB130603B_kNe_candidate, tanvir13_GRB130603B_kNe_candidate} may indeed be due to emission from heated dust, which is an alternative explanation to the kilonova model. The specific luminosity of the excessive emission is $8\times10^{26}\rm\, erg\,s^{-1}\,Hz^{-1}$ at rest-frame wavelength $\lambda_{\rm obs}\simeq 1.2\um$, which is slightly shorter than the peak of the emission spectrum for dust temperature $T_{\rm sub}\simeq 2200\rm\, K$. From eq. (\ref{eq:on_axis_luminosity}), we see that the dust echo may be consistent with the excess, as long as (1) the isotropic equivalent UV energy from the jet is $E_{\rm UV}\sim\,$a few$\times10^{50}\rm\, erg$ and (2) an optically thick dust layer exists at a distance of $r_{\rm sub}\sim 3\rm\, pc$. The first requirement can be satisfied if the jet energy is $E_{\rm j}\sim 3\times10^{52}\rm\, erg$ and the CSM density at a distance of $0.1\rm\, pc$ is higher than $0.3\rm\, cm^{-3}$ (cf. Appendix \ref{sec:reverse_shock}). An optically thick ($A_{\rm V}\sim 1\rm\, mag$) dust layer of thickness $\sim 3\rm\,pc$ means that the hydrogen number density is $n_{\rm H}\sim 300\rm\, cm^{-3}$. This is possible if the source is embedded in or near a molecular cloud. In fact, the steep spectrum of the optical afterglow from this event indeed showed LOS dust extinction of $A_{\rm V}\simeq 1\rm\, mag$ in the star-forming host galaxy \citep{deUgartePostigo14_GRB130603B_host, fong14_GRB130603b_afterglow}.

\begin{figure*}
\centering
\includegraphics[width=0.5\textwidth]{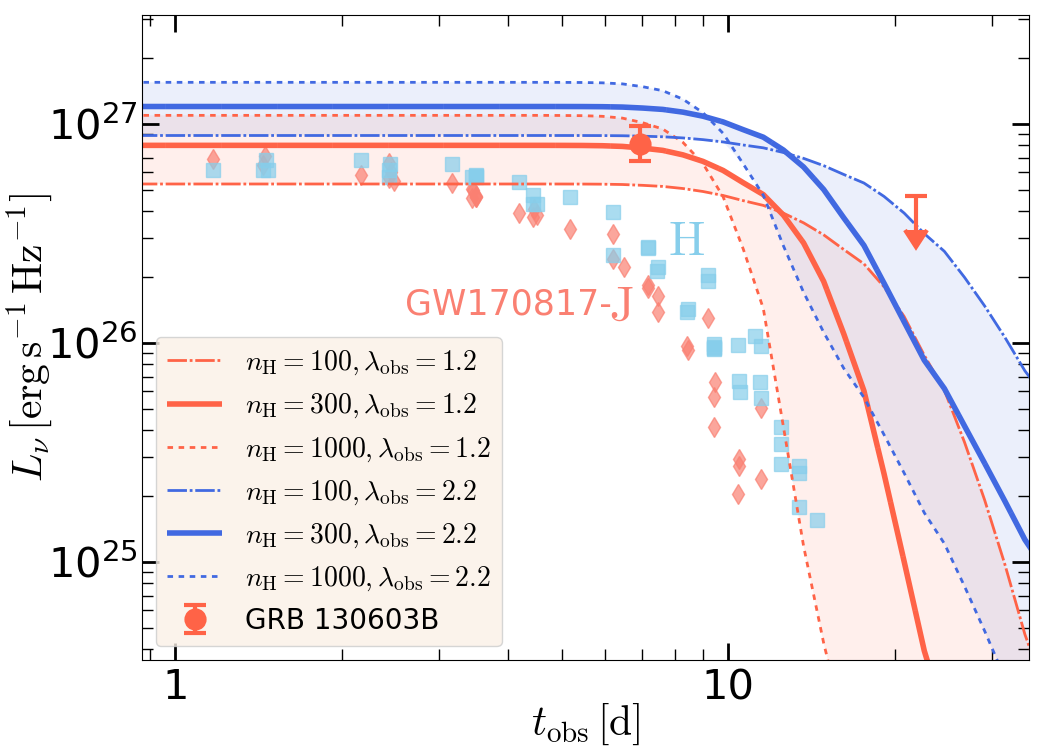}
\caption{Dust echo lightcurves viewed from an on-axis LOS ($\theta_{\rm obs}=0$) at host-galaxy rest-frame wavelength $\lambda_{\rm obs}=1.2\um$ (red) and $2.2\um$ (blue lines), and for different hydrogen number densities $n_{\rm H} = 100\rm\, cm^{-3}$ (dash-dotted lines), $300\rm\, cm^{-3}$ (solid lines), and $1000\rm\, cm^{-3}$ (dotted lines). The red circle and arrow denotes the H-band (rest-frame $\lambda_{\rm obs}\simeq 1.2\um$) flux excess at $t_{\rm obs}=6.9\rm\,d$ and H-band upper limit at $t_{\rm obs}=21.8\rm\,d$ from GRB 130603B \citep{berger13_GRB130603B_kNe_candidate, tanvir13_GRB130603B_kNe_candidate}, respectively. For all cases, we have fixed the jet opening angle $\theta_{\rm j}=4^{\rm o}$, and dust-to-gas ratio parameter $n_0/n_{\rm H} = 1.45\times10^{-15}$ (which is calibrated by the extinction law of a typical Galactic LOS). The emission from the jet is taken to be a square pulse with isotropic equivalent UV-optical luminosity $L_{\rm UV} = 10^{48}\rm\, erg\, s^{-1}$ and duration $t_{\rm UV}=500\rm\, s$. For comparison, we also show the J-band ($\lambda_{\rm obs}\simeq1.2\um$, red diamonds) and H-band ($\lambda_{\rm obs}\simeq2.2\um$, blue squares) lightcurves of the kilonova associated with GW170817 \citep{villar17_kilonova_data_compiled}. 
}
\label{fig:kNe-compare}
\end{figure*}

As a concrete example, we consider the UV-optical emission from the jet to be a square pulse with isotropic equivalent UV-optical luminosity $L_{\rm UV} = 10^{48}\rm\, erg\, s^{-1}$ and duration $t_{\rm UV}=500\rm\, s$. The dust echo lightcurves for an on-axis observer are shown in Fig. \ref{fig:kNe-compare}, for three different hydrogen number densities of $n_{\rm H}=100$, 300, $1000\rm\, cm^{-3}$. We find that the H-band excess in GRB 130603B is consistent with the dust echo model. We note that the H-band excess can also be reasonably explained by a kilonova from lanthanide-rich ejecta of mass $0.05$--$0.1\,M_\odot$ and expansion speed $\sim0.2c$, as modeled by the original authors \citep{berger13_GRB130603B_kNe_candidate, tanvir13_GRB130603B_kNe_candidate}. A near-IR spectrum can be used to distinguish between these two scenarios: a kilonova spectrum has features due to atomic transitions \citep[e.g.,][]{watson19_kNe_spectrum} whereas the spectrum from dust emission is expected to be featureless. The high grain temperature means that a large number of vibrational modes are excited and the energy levels occupied are those with very large quantum numbers above the ground state \citep{draine01_heating_of_small_grains}, so the emission spectrum is continuous. The silicate feature near $10\rm\, \um$, due to O-Si-O strentching mode which has a large dipole matrix element, may stand out above the continuum when the grain is hotter than the excitation temperature of $\sim$$200\rm\, K$. We conclude that future kilonova searches in short GRB afterglow photometric data should be aware of the possible contamination from dust echo, especially when the optical-IR afterglow spectrum shows evidence for strong dust extinction ($A_{\rm V}\gtrsim 0.5\rm\,mag$) or when the X-ray spectrum shows large neutral-hydrogen column density ($N_{\rm H}\gtrsim 10^{21}\rm\, cm^{-3}$).

\section{Dust Echo Model}\label{sec:model}
In this section, we construct a detailed model to calculate the lightcurves of dust echoes as well as the intensity distribution as viewed from an arbitrary LOS. 

To predict the emission from the UV-heated dust, we need a model for the distribution of grain sizes and how they interact with (i.e., absorb, scatter, and emit) radiation from the infrared to the UV bands. In the last few decades, significant progress has been made in understanding the size distribution, composition, and various frequency-dependent features in the dielectric functions, based on measurements of the extinction curve, spectra of reflection nebulae, polarization of starlight, near-infrared emission features, thermal emission from cold dust at longer wavelengths, as well as \textit{in situ} measurement of interstellar dust in the Solar System \citep[see][and refs therein]{Draine03_araa, draine11_book}. However, since the interstellar medium is dynamic, multi-phase and highly complex, even the ``best-fit" models are only crude approximations to the real physical situations. In this work, we adopt the following simplified model that captures the main physical processes relevant to our discussion.

We consider the \citet[][hereafter MRN]{mathis77_grain_size_distribution} distribution of grain sizes,
\begin{equation}\label{eq:dnda}
    {\d n_{\rm g}\over \d \aum} = n_0 a_{\rm \mu m}^{-3.5},\ a_{\rm min} < a_{\rm \mu m} < a_{\rm max},
\end{equation}
where $n_0$ is the normalization in units of $\rm\, cm^{-3}$, $a$ is the radius of an equal-volume sphere for each grain in units of $\rm \mu m$, and we take minimum size $a_{\rm min}=0.01\,\um$ and maximum size $a_{\rm max}=0.3\,\um$. Much smaller grains $a< 0.01\,\um$ (e.g., polycyclic aromatic hydrocarbons or PAHs) are optically thin, evaporate quickly and hence do not contribute significantly to dust echo. Much larger grains $a>0.3\,\um$ (if they exist) make a minor contribution to the extinction of optical and UV radiation from the source. In the long wavelength limit $\lambda\gg 2\pi a$, the extinction is dominated by absorption, and the wavelength dependence of the absorption cross-section is given by $C_{\rm abs,\lambda}\propto a^3\lambda^{-2}$ under the electric dipole approximation \citep[eq. 22.27 in][]{draine11_book}. At shorter wavelengths\footnote{In the special case of a spherical grain, the absorption/scattering cross-sections can be computed by the Mie theory, which has been widely used to model interstellar dust \citep[e.g.,][]{draine84_Qext_extinction_curve}. However, physical dust particles are non-spherical and are much more difficult to model accurately.}, both absorption and scattering may be important, and the total extinction cross-section $C_{\rm ext, \lambda}$ asymptotically approaches twice the geometric area of $\pi a^2$ in the limit $\lambda \ll 2\pi a$, as long as the grain is optically thick. This motivates us to consider the following model for the efficiency factors for absorption and extinction
\begin{equation}\label{eq:Qabs_lambda}
    Q_{\rm abs,\lambda} \equiv {C_{\rm abs,\lambda}\over \pi a^2} \simeq {1\over 1 + (\lambda/\lambda_0)^2 \aum^{-1}},\ 
    Q_{\rm ext,\lambda} \equiv {C_{\rm ext,\lambda}\over \pi a^2} \simeq {1\over 1/2 + (\lambda/\lambda_0)^2 \aum^{-1}},
\end{equation}
where $\lambda_0$ is a critical wavelength (typically a few to 10 $\rm \mu m$) that depends on grain composition and internal structures. This simple model captures the broad-brush picture of dust absorption/emission and makes the (otherwise complicated) integrals over the grain size distribution analytically tractable. We adopt $\lambda_0 \simeq 2\, \um$ which gives Planck-averaged absorption efficiency (see eq. \ref{eq:QabsP}) in between graphite and silicate models by \citet{laor93_Qabs} in the temperature range (500 to $2500\rm\, K$) of interest for dust echo.

The optical depth due to dust extinction can be integrated analytically
\begin{equation}
    {\tau_{\rm d,\lambda} \over N_{\rm H}} = {\pi \mr{\mu m}^2 \over n_{\rm H}}\int_{a_{\rm min}}^{a_{\rm max}} \d \aum {\d n_{\rm g}\over \d \aum}  \aum^2 Q_{\rm ext,\lambda} =
    {2\sqrt{2}\pi \mr{\mu m}^2 n_0 \over n_{\rm H}} {\lambda_0 \over \lambda} \left(\mr{atan}\sqrt{0.5x_{\rm max}} - \mr{atan}\sqrt{0.5x_{\rm min}}\right),
\end{equation}
where $x_{\rm min/max} = a_{\rm min/max}(\lambda_0/\lambda)^2$, $n_{\rm H}$ is the H number density (including atomic, ionic and molecular forms), and $N_{\rm H}$ is the H column density. For $\lambda_0$ of a few $\um$, this roughly reproduces the observed extinction curve of $A_\lambda/\mr{mag} = 1.086\tau_{\rm d,\lambda} \propto \lambda^{-1}$ in the optical and UV for a typical Galactic line of sight with $R_{\rm V}=3.1$ \citep{cardelli89_extinction_law, fitzpatrick99_extinction_law}.
We normalize the extinction curve by V-band extinction $A_{\rm V}/N_{\rm H}\simeq 5.3 \times 10^{-22} \rm\, mag\, cm^2$ at $\lambda\simeq 0.55\um$. This gives $n_0/n_{\rm H}\simeq 1.45\times10^{-15}$ and hence
\begin{equation}\label{eq:extinction_law}
    {\tau_{d,\lambda}}\simeq 0.49 {A_{\rm V}/\mr{mag}\over \lambda_\um} \left(\mr{atan}\sqrt{0.5x_{\rm max}} - \mr{atan}\sqrt{0.5x_{\rm min}}\right).
\end{equation}
In this way, our simplified dust model (eqs. \ref{eq:dnda} and \ref{eq:Qabs_lambda}) gives qualitatively similar extinction curve as the more sophisticated model by e.g., \citet{weingartner01_grain_size_distribution}. A comparison is shown in Fig. \ref{fig:extinction} in the Appendix. We note that for $A_{\rm V}\sim 1\rm\, mag$, the gas-dust column is highly optically thick for UV photons, which are strongly attenuated by a small fraction of the column density.

At a given time, the specific luminosity $L_\nu$ is given by afterglow emission from the jet-driven external shocks, and the heating rate for a grain of size $a$ at a distance $r$ from the source is given by
\begin{equation}\label{eq:heating_rate}
    \dot{q}_{\rm h} = \int^{\nu_{\rm max}} \d \nu {L_\nu \mr{e}^{-\tau_\lambda} \over 4\pi r^2} \pi a^2 Q_{\rm abs,\lambda},
\end{equation}
where we consider maximum photon energy of $h\nu_{\rm max}=50\rm\, eV$ (or $\lambda_{\rm min}\simeq 0.025\um$), because higher energy photons have smaller absorption cross-sections (see Fig. \ref{fig:extinction}). The cooling rate due to thermal emission at grain temperature $T$ is given by (using Kirkhoff's theorem)
\begin{equation}
    \dot{q}_{\rm c} = 4\pi a^2\int Q_{\rm abs,\lambda} \,\pi B_\nu \d \nu = 4\pi a^2 \QabsP \sigma_{\rm SB} T^4,
\end{equation}
where $\sigma_{\rm SB}$ is the Stefan-Boltzmann constant, $B_\nu(T)$ is the Planck function, and the Planck-averaged absorption efficiency is defined as
\begin{equation}\label{eq:QabsP}
    \QabsP \equiv {\pi\over \sigma_{\rm SB}T^4}{\int Q_{\rm abs,\lambda} B_\nu \d \nu}.
\end{equation}
For absorption efficiency in eq. (\ref{eq:Qabs_lambda}), we find that the Planck-averaged $\QabsP$ can be approximated by
\begin{equation}\label{eq:QabsP_approx}
    \QabsP\simeq  {1\over 1 + 3.5 (T/1000\mr{\,K})^{-2} \aum^{-1}},
\end{equation}
which is in between the graphite and silicate models of \citet{laor93_Qabs} and is also similar to the approximation adopted by \citet[][their eq. 13]{waxman00_GRB_dust_echo}. A comparison is shown in Fig. \ref{fig:QabsP} in the Appendix. Therefore, the dust temperature $T(a, r)$ at a given time\footnote{This is the retarded time, i.e., the time since the arrival of the first causal signal from the center of explosion.} $t$ is given by the balance between heating and cooling,
\begin{equation}\label{eq:equilibrium}
    \dot{q}_{\rm h} = \dot{q}_{\rm c} \ \ \Rightarrow\ \ T(a, r, t).
\end{equation}
Using the approximated Planck-averaged $\QabsP$, the energy balance can be written into a simple cubic equation whose solution is given by
\begin{equation}
    {T\over 1000\mr{\,K}} =\left[\left(2y^2\over 3\xi\right)^{1\over 3} + \left(\xi y \over 18\right)^{1\over 3}\right]^{1/2}, \ y = {\dot{q}_{\rm h} [\mr{cgs}]\over 7.13\aum^2}, \ \xi = \left[\left(31.5\over \aum\right)^2 - 12y\right]^{1/2} + {31.5\over \aum}.
\end{equation}
The above expression breaks down at extremely high heating rate when $y> (31.5/\aum)^2/12$, and in that situation, the dust temperature is higher than $3240\aum^{-1/2}\rm\, K$ and the physical consequence is that the grain evaporates almost immediately. In the limit of $y\ll \xi^2/12$, the second term in the temperature expression dominates, and we have $\xi\simeq 62/\aum$ and $y\simeq 3.5L_{48}/r_{19}^2$ (ignoring attenuation), so a rough estimate of the dust temperature is given by $T\simeq 2.2\times10^3\mr{\,K}\, L_{48}^{1/6} r_{19}^{-1/3} a_{0.1\um}^{-1/6}$.

The grain sublimation rate at a given temperature is determined by the balance between atoms evaporating from the surface and those returning from the gas phase, which are poorly known (given our limited knowledge about grain composition, surface properties, and shape). We adopt the (crude) estimate of the sublimation rate given by \citet{waxman00_GRB_dust_echo},
\begin{equation}
    {\d a\over \d t}\sim 2\times10^7 \mr{cm\,s^{-1}}\, \mr{exp}(-7\times10^4\mr{K}/T).
\end{equation}
Then, we estimate the sublimation temperature $T_{\rm sub}$ by equaling the characteristic survival time $t_{\rm surv}=a/|\d a/\d t|$ to the current time $t$ since the arrival of the first causal signal, and this gives
\begin{equation}
    T_{\rm sub} \simeq 2.33\times10^3\mr{K} \left[1 - 0.033\,\mr{ln}{t/100\mr{\,s}\over {\aum}}\right].
\end{equation}
We further define a sublimation radius $r_{\rm sub}(a, t)$ by
\begin{equation}
    T(a, r_{\rm sub}, t) = T_{\rm sub},
\end{equation}
where $T(a, r, t)$ is given by eq. (\ref{eq:equilibrium}). 
For simplicity, we assume that all grains with temperature $T>T_{\rm sub}$ evaporate instantaneously. This is a good approximation because the survival time is a very sensitive function of temperature (a few percent variation in $T$ gives an order unity change in $t_{\rm surv}$), and it can be shown that the evaporating layer with only has radial thickness of a few percent of the sublimation radius \citep[e.g.,][]{lu16_TDE_echo, sun20_TDE_echo_model}. 

Finally, the attenuation of the source flux, $\tau_\lambda=\tau_{\rm d,\lambda} + \tau_{\rm g,\lambda}$, is dominated by dust extinction in regions where the gas is fully ionized, but when gas is largely neutral, bound-free absorption dominates the opacity for ionizing photons above $13.6\rm\, eV$ and dust extinction dominates for lower energy photons.
Neutral H and He have very large bound-free absorption cross-sections for photons below about $100\rm\, eV$, and they are ionized up to a radius $r_{\rm ion}(t)$, which is given by
\begin{equation}\label{eq:ionization_radius}
    \int_0^{r_{\rm ion}(t)} 4\pi r^2 n_{\rm H} \d r \simeq \int_0^{t}\d t \int_{13.6\rm eV}^{100\rm eV} {L_\nu\over h\nu} \d \nu,
\end{equation}
where the right-hand side is the cumulative number of ionizing photons. Note that gas recombination can be ignored because of the very long timescale. We assume that gas opacity is negligible at radius $r < r_{\rm ion}$ and that there are no ionizing photons ($>13.6\rm\, eV$) beyond radius $r_{\rm ion}$. Note that in (\ref{eq:ionization_radius}), we have neglected dust attenuation at radii $r<r_{\rm ion}$, which is only important when most of the source radiation is attenuated by dust instead of being spent on gas ionization. Therefore, this does not affect our goal of calculating the dust heating rate in eq. (\ref{eq:heating_rate}) by accounting for the modification of the source spectrum due to gas ionization.

On the other hand, the dust extinction optical depth is given by
\begin{equation}\label{eq:extinction_with_sublimation}
\begin{split}
    \tau_{\rm d,\lambda}(<r) &\simeq \pi \mr{\mu m}^2 \int_0^{r} \d r \int_{a_{\rm sub}(r)}^{a_{\rm max}} \d \aum {\d n_{\rm g}\over \d \aum}  \aum^2 Q_{\rm ext,\lambda}\\
    &= {2\sqrt{2}\pi \um^2 n_0 \lambda_0 \over n_H \lambda} \int \d r\, n_{\rm H} \left(\mr{atan}\sqrt{0.5x_{\rm max}} - \mr{atan}\sqrt{0.5x_{\rm sub}(r)}\right),\ x_{\rm max/sub} = a_{\rm max/sub} (\lambda_0/\lambda)^2,
\end{split}
\end{equation}
where we have taken into account that grains below a critical size (if $<a_{\rm min}$) has already evaporated, and the critical size $a_{\rm sub}(r, t)$ is given by $T(a_{\rm sub}, r, t)=T_{\rm sub}$. At sufficiently large distances from the source where even the smallest grain survives the heating, we take $a_{\rm sub} = a_{\rm min}$.

The system of equations above needs to be solved iteratively, because dust sublimation affects the attenuation of the source flux, which in turn affects dust sublimation. At each time $t$, we carry out the first iteration to find $a_{\rm sub}(r, t)$ at each radius $r$ when ignoring dust attenuation at small radii $<r$. Then, the result from the first iteration is used in the next iteration which takes into account dust attenuation by eq. (\ref{eq:extinction_with_sublimation}). This modifies $a_{\rm sub}(r, t)$, which is used in the next iteration until convergence is achieved. In each successive iteration, we expect $a_{\rm sub}(r, t)$ to be reduced\footnote{Smaller grains are hotter and hence easier to evaporate, because they are less efficient at cooling.} and there will be slightly more dust attenuation. The whole process is done at each time step $t$ since the arrival of causal signal to each of radial shells at different $r$'s. We find that convergence is achieved within 5 iterations. Note that dust sublimation is irreversible, so $a_{\rm sub}(r, t)$ is a non-decreasing function of time, and the grain size distribution at each radius is frozen after the peak luminosity of the source light curve.

\begin{figure*}
\centering
\includegraphics[width=0.45\textwidth]{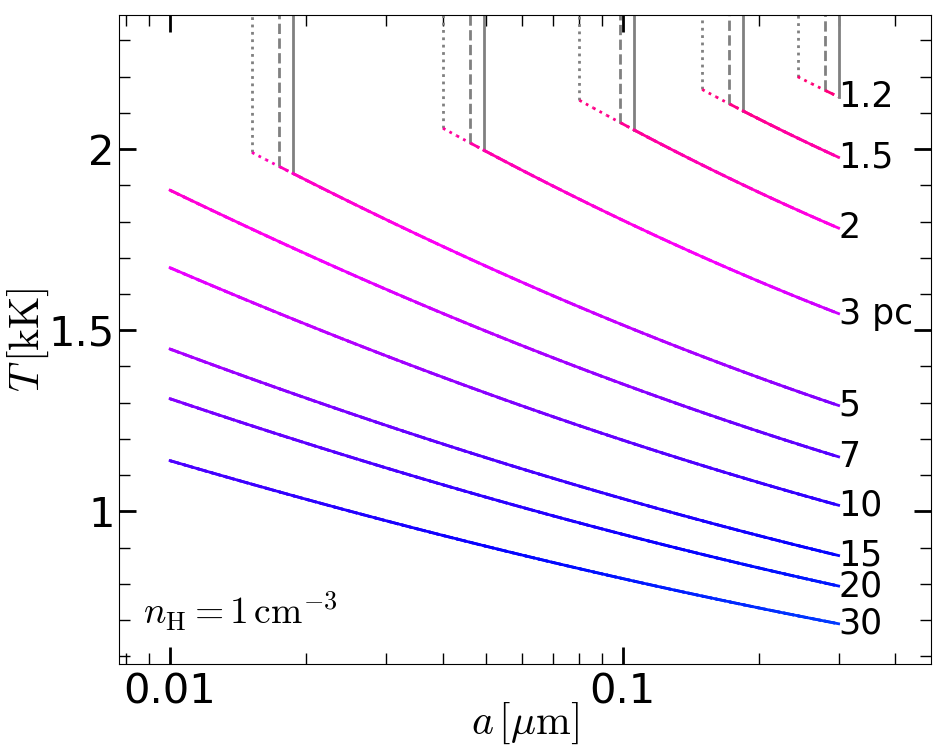}
\includegraphics[width=0.45\textwidth]{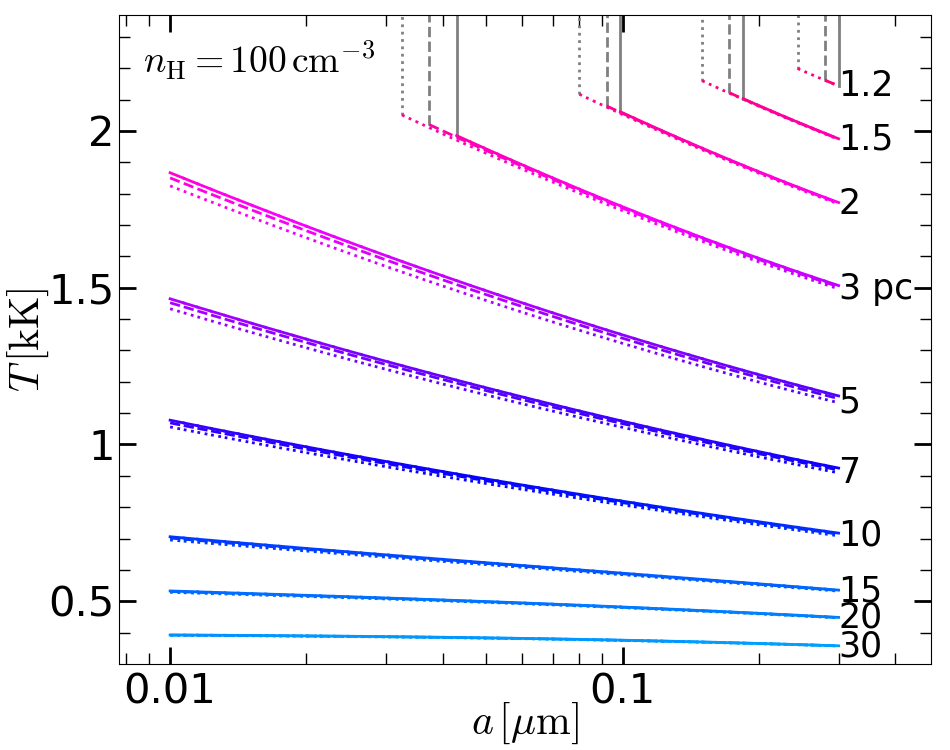}
\caption{Dust temperatures (in units of kilo-Kelvin) as a function of grain size $a$ (horizontal axis) and radius $r$ from the center (marked at the end of each line in units of pc). The \textit{left panel} is for uniform gas density $n_{\rm H}=1\rm\,cm^{-3}$ and the \textit{right panel} is for much higher gas density of $100\rm\, cm^{-3}$ (the results for $n_{\rm H}\leq 10\rm\, cm^{-3}$ at $r\lesssim 30\rm\,pc$ are very similar to the $1\rm\, cm^{-3}$ case). The jet emission is taken to be a square pulse with isotropic equivalent UV-optical luminosity $L_{\rm UV} = 3\times 10^{47}\rm\, erg\, s^{-1}$ and duration $t_{\rm UV}=300\rm\, s$ (the same as in Fig. \ref{fig:echo_theobs20}). At each radius, we show the temperature vs. grain sizes at three different times $t = 100\rm\,s$ (dotted), $200\rm\,s$ (dashed), $300\rm\, s$ (solid lines). The initial grain size distribution has minimum $a_{\rm min}=0.01\um$ and maximum $a_{\rm max}=0.3\um$. We find that smaller grains are hotter and hence sublimate earlier and farther from the source. The vertical grey lines indicate the critical grain size $a_{\rm sub}(r, t)$ below which all grains at radius $r$ and time $t$ have sublimated. At sufficiently large radii ($r\gtrsim 7\rm\, pc$ on the left panel and $r\gtrsim 5\rm\, pc$ on the right), all grains considered in our model survive the entire radiative heating event, and hence all three lines of different styles overlap. On the right panel, due to high gas/dust density, the outer radii ($r\gtrsim 10\rm\, pc$) are significantly shielded by the surviving dust at smaller radii, so the dust far away from the source are only weakly heated to $T\lesssim 1\rm\, kK$ and their emission is in the mid-IR.
}
\label{fig:Td}
\end{figure*}

The final result of the above calculation is that we obtain the time evolution of the temperatures of (surviving) dust grains of all sizes $a$ at all radii $r$, $T(a, r, t)$, as shown in Fig. \ref{fig:Td} for two representative cases. This is then used to calculate the light curve from dust emission. So far, everything has been calculated under spherical symmetry, which is only broken by the fact that the afterglow emission from the jet is beamed into a narrow solid angle and that the observer's line of sight is misaligned with the jet axis. The emissivity\footnote{Here the emissivity is defined as the power radiated per unit volume (containing a large number of grains) per solid angle per frequency, in units of $\rm erg\,s^{-1}\, cm^{-3}\, sr^{-1}\, Hz^{-1}$.} from the heated dust at space-time position $(\vec{r}, t)$ is given by
\begin{equation}
\begin{split}
    j_{\rm d,\nu}(r, t) &={\um^2\over 4\pi} \int_{a_{\rm sub}(r)}^{a_{\rm max}} \d \aum {\d n_{\rm g}\over \d \aum} 4\pi \aum^2 Q_{\rm abs,\lambda}\, \pi B_\nu(T(a, r, t))\\
    & = {2\pi h\nu\over \lambda_\um^2} n_0(r) \int_{a_{\rm sub}(r)}^{a_{\rm max}} \d \aum {\aum^{-0.5}\over \aum + (\lambda/\lambda_0)^2} {1\over \mr{e}^{h\nu/kT} - 1}.
\end{split}
\end{equation}
where
we have ignored any further attenuation of the dust infrared emission. This emission will arrive at the observer after a time delay (since the arrival of the first causal signal from the explosion)
\begin{equation}
    t_{\rm obs} = t + {r\over c} - {\vec{r}\cdot \hat{e}_{\rm LOS} \over c},
\end{equation}
where $\hat{e}_{\rm LOS} = (x= \sin\theta_{\rm obs}, y=0, z = \cos\theta_{\rm obs})$ is a unit vector pointing from the center of explosion to the observer. All vectors are expressed in Cartesian components where the $\hat{z}$ axis is aligned with the jet axis and the observer's LOS is in the $x$-$z$ plane. The angle between the jet axis and the LOS is denoted as $\theta_{\rm obs}\in (0, \pi/2)$. We divide the entire volume illuminated by the source radiation into (logarithmic) radial shells. At a given observer's time $t_{\rm obs}$, only a narrow stripe on each radial shell satisfies the time-delay constraint and the angular extent of this stripe (for a fixed $r$) is given by
\begin{equation}\label{eq:light_travel_delay}
    1 - \tmu = {c\over r}(t_{\rm obs} - t), \ \ 0 < t < t_{\rm max},
\end{equation}
where $t_{\rm max}$ is the maximum duration of the source emission, $\tmu = \cos\tilde{\theta} = \hat{r}\cdot \hat{e}_{\rm LOS}$ the cosine of the angle between the radial vector and the LOS. From time-delay constraint alone, the minimum $\tmu_{\rm min}$ (or maximum $\tmu_{\rm max}$) corresponds to $t=0$ (or $t = t_{\rm max}$). Another requirement is that the stripe must be within the beaming cone of the jet radiation, which has an angular size of $\theta_{\rm j}$. The total isotropic equivalent specific luminosity of the dust echo at time $t_{\rm obs}$ can be written as
\begin{equation}\label{eq:echo_luminosity}
    L_{\rm d,\nu}(t_{\rm obs}) = 4\pi \int \d r\, r^2 \int_{\tmu_{\rm min}}^{\tmu_{\rm max}} \d\tmu\, j_{\rm d,\nu}(r, t(\tilde{\theta})) \, 2\tilde{\phi}_{\rm max},
\end{equation}
where $\tilde{\phi}_{\rm max}(\tmu)$ is the angle between $\hat{r}_{\rm j} - \tmu \hat{e}_{\rm LOS}$ and the $x$-$z$ plane
\begin{equation}\label{eq:tphi_max}
    \sin \tilde{\phi}_{\rm max} = {(\hat{r}_{\rm j} - \tmu \hat{e}_{\rm LOS}) \cdot{\hat{y}} \over |\hat{r}_{\rm j} - \tmu \hat{e}_{\rm LOS}|} = {\sin\theta_{\rm j} \sin\phi\over \sin \tilde{\theta}}
\end{equation}
and $\hat{r}_{\rm j}=(x=\sin\theta_{\rm j}\cos\phi, y=\sin\theta_{\rm j}\sin\phi, z=\cos\theta_{\rm j})$ is the intersection between the edge of the jet cone and the cone at an angle $\tilde{\theta}$ from the LOS, i.e.
\begin{equation}\label{eq:jet_cone_edge}
    \tmu = \sin\theta_{\rm obs} \sin\theta_{\rm j} \cos\phi + \cos\theta_{\rm obs}\cos\theta_{\rm j}.
\end{equation}
For each $\tmu$, we solve for $\phi$ from eq. (\ref{eq:jet_cone_edge}) and then obtain $\tilde{\phi}_{\rm max}$ from eq. (\ref{eq:tphi_max}).
Note that when the observer's LOS is within the jet beaming cone, i.e. $\theta_{\rm obs}<\theta_{\rm j}$, we have $\tilde{\phi}_{\rm max} = \pi$ for $\tilde{\theta}<\theta_{\rm j} - \theta_{\rm obs}$ because the entire azimuth (wrt. the LOS) is illuminated by the jet radiation. The requirement that the emitting region is within the beaming cone of the jet radiation means that $\max(0,\ \theta_{\rm obs} - \theta_{\rm j})<\tilde{\theta}<\theta_{\rm obs} +\theta_{\rm j}$, and this puts further constraints on $\tmu_{\rm min/max}$ besides the time-delay requirement. Another useful constraint to expedite the radial integral in eq. (\ref{eq:echo_luminosity}) is that, at a given observer's time $t_{\rm obs}$, only the layers above a minimum radius $r_{\rm min}$ may contribute to the observed flux, and $r_{\rm min} = \max[0, \ c(t_{\rm obs}-t_{\rm max})]/[1 - \cos(\theta_{\rm obs} + \theta_{\rm j})]$. When the observer's LOS is outside the jet beaming cone $(\theta_{\rm obs}>\theta_{\rm j})$, there is also a maximum radius $r_{\rm max} = ct_{\rm obs}/[1 - \cos(\theta_{\rm obs} - \theta_{\rm j})]$, beyond which any light echo has not arrived at the observer yet.

Finally, we calculate the intensity distribution of the dust emission and the position of the flux centroid.
We consider another frame of Cartesian coordinates $(\tx, \ty, \tz)$ with its origin at the center of explosion, the $\hat{\tz}$ axis pointing towards the observer, the base vector $\hat{\tx}$ inside the $x$-$z$ plane of the old coordinate system used to describe the jet, and the third base vector $\hat{\ty}$ is anti-parallel to $\hat{y}$. Note that since $\hat{\tz}\cdot \hat{z} = \cos\theta_{\rm obs} > 0$, we have $\hat{\tx}\cdot\hat{z} = \sin\theta_{\rm obs}$. Under this set up, the position of a volume element in spherical coordinates $(r,\tmu=\cos\ttheta, \tphi)$ can be expressed in Cartesian components $(\tx = r\sin\ttheta \cos\tphi, \ty=r\sin\ttheta\sin\tphi, r\tz=\cos\ttheta)$ or more conveniently in cylindrical components $(\trho=r\sin\ttheta, \tphi = \tphi, \tz = r\cos\ttheta)$.

We are interested in the intensity distribution $I_{\rm d,\nu}(t_{\rm obs}, \trho, \tphi)$ on the sky (with the sky positions in cylindrical coordinates) as a function of the observer's time $t_{\rm obs}$. Since we ignore the attenuation of dust emission, the intensity field is given by the linear superposition of many radial spherical shells, each of which has emissivity $j_{\rm d,\nu}(r, t)$.
The intensity contribution from each radial shell from $r$ to $r+\d r$ is
\begin{equation}
\d I_{\rm d, \nu} =
\begin{cases}
 j_{\rm d,\nu}(t, r) \tmu^{-1} \d r,\ \ \mbox{if the volume element is within the jet beam},\\
0,\ \ \mr{otherwise}.
\end{cases}
\end{equation}
Note that the local time $t$ is given by the light-travel time condition of eq. (\ref{eq:light_travel_delay}).
Then, we integrate over all radial shells to obtain the total intensity distribution at a given observer's time $t_{\rm obs}$, $I_{\rm d,\nu}(t_{\rm obs}, \trho, \tphi) = \int \d I_{\rm d,\nu}$. For the case where the intensity image is symmetric along the $\ty$ direction, the projected position of the flux centroid is given by (on the $\tx$ axis)
\begin{equation}
    \langle\tx\rangle_{I_{\rm d,\nu}}(t_{\rm obs}) = {\int\int\tx I_{\rm d, \nu}\d \tx \d \ty \over \int\int I_{\rm d, \nu}\d \tx \d \ty} = {\int\int \trho\cos\tphi\, I_{\rm d, \nu}\trho\, \d \trho\, \d \tphi \over \int\int I_{\rm d, \nu}\trho\, \d \trho\, \d \tphi}. 
\end{equation}
If we only want to know the position of the flux centroid, then a simpler way is to directly compute the weighted mean of $<\trho\cos\tphi>_{\tphi} = r\sqrt{1-\tmu^2}\, \tphi_{\rm max}^{-1} \sin\tphi_{\rm max}$ in the luminosity integral of eq. (\ref{eq:echo_luminosity}).




\section{Effects of X-ray photons}\label{sec:X-ray}

Although this paper focuses on the dust echo of the UV-optical emission from the reverse shock, one should note that absorption of soft X-rays $\lesssim \rm a \,few\, keV$ near the L- and K-edges of constituent atoms also contributes to dust heating \citep{fruchter01_xray_destruction}, since the energy goes into the primary photoelectrons as well as Auger electrons that are largely trapped by sufficiently large grains \citep[][and only a small fraction of energy is lost in fluorescence photons, e.g., K$\alpha$]{weingartner06_grain_charging}. The prompt emission of short-hard GRBs has $\nu L_\nu$ peak energy $\gtrsim 300\rm\, keV$ and the typical low-energy photon index is $-1$ as routinely measured by \textit{Swift} BAT down to about 20 keV \citep{kumar15_GRB_review}, so only a small fraction ($\sim 1\%$) of the prompt-emission energy is emitted in the soft X-ray band near a few keV. A potentially stronger source of X-ray emission is the external forward shock. In the most optimistic case where most of the shock-accelerated electrons emit at $ \sim \rm a \,few\, keV$ and their cooling time is shorter than the dynamical expansion time (i.e., in the fast cooling regime), a fraction $\epsilon_{\rm e}\sim 10\%$ of the jet kinetic energy can be emitted in the soft X-ray band over the jet deceleration timescale. If this is the case, then the X-ray energy can be higher than $E_{\rm UV}$ by an order of magnitude. However, the dust absorption cross-section in the soft X-ray band is about an order of magnitude smaller than that in the UV \citep{Draine03_araa}, so the dust heating provided by soft X-rays may be comparable to the UV-optical heating and will make the dust echo slightly brighter than our prediction.

It has also been pointed out that absorption of hard X-rays $\gtrsim \rm a \,few\, keV$ generates energetic photoelectrons as well as Auger electrons that are capable of escaping the grain \citep{weingartner06_grain_charging}, and hence the grain will be strongly positively charged. As the grain charge builds up, it becomes more and more difficult for electrons to escape to infinity and hence higher energy X-ray photons are required to further increase the ionization. For a spherical grain with charge number $Z_{\rm g}$, the Coulomb electric field $E_{\rm Coul}\sim Z_{\rm g}e/a^2\sim 1.4\mr{\,V/\AA}\, (Z_{\rm g}/10^5) a_{\rm 0.1\um}^{-2}$. This means that if the grain charge number exceeds a critical value $Z_{\rm g,max}\sim 10^5 a_{\rm 0.1\um}^2$ (or if the grain surface potential exceeds $Z_{\rm g,max}e/a\sim 1.4\rm\, kV\, a_{\rm 0.1\um}$), the Coulomb field is capable of breaking molecular bonds of typical strength $\sim 1\rm\, eV$. Thus, the grain will likely be broken apart in a process called ``Coulomb explosion'' \citep{waxman00_GRB_dust_echo, fruchter01_xray_destruction}. The total number of electrons per grain is $N_{\rm e}\sim 3\times 10^{9} a_{\rm 0.1\um}^{3}$, and the effective absorption cross-section \textit{per electron} near $5\rm\, keV$ is of the order\footnote{To calculate $\bar{\sigma}_{\rm e}$, one has to take into account the photoelectric absorption probability and the escape probability under a large positive potential of $\sim$kV at the grain surface. We estimate $\bar{\sigma}_{\rm e}$ by dust absorption cross-section of $10^{-24}\rm\, cm^2$ per H near $5\rm\, keV$ \citep{draine03_X-ray_scattering} divided by the dust-to-gas mass ratio of $10^{-2}$ for solar metallicity.} $\bar{\sigma}_{\rm e}\sim 10^{-22}\rm\, cm^{-2}$, so Coulomb explosions require an X-ray photon fluence of $F_{\rm X}\gtrsim 3\times10^{17}\,\mr{cm}^{-2} a_{\rm 0.1\um}^{-1}$ or isotropic equivalent energy of $E_{\rm X}(\sim 5\rm keV)\gtrsim 10^{47}\mr{\,erg}\, a_{\rm 0.1\um}^{-1} (r/1\mr{\,pc})^2$ at a distance of $r$ from the center of explosion. Thus, it has been argued that Coulomb explosions might occur up to tens of parsecs from the source in the jet beaming cone \citep{fruchter01_xray_destruction}.

However, it is unclear whether a large electrostatic stress sufficient to cause Coulomb explosions can build up in the main body of the grain. If a grain is sufficiently conducting and has an irregular shape, charge concentrations near sharp corners will lead to much stronger local Coulomb fields, which may break the molecular bonds, ionize the atoms and then expel the ions from the tip of the corners --- this process is called ion-field emission or field evaporation \citep{brandon1968field_evaporation}. Ion-field emission as a way of limiting the total grain charge has been discussed by \citet{draine79_grain_in_hot_gas, draine02_molecular_gas_absorption}. 
For instance, a sharp substructure with curvature radius $\tilde{a}\ll a$ ($a$ being the size of the main grain body and $\tilde a$ may be as small as the atomic size) will contain charge number $\tilde{Z}_{\rm g}\sim (\tilde{a}/a) Z_{\rm g}$, based on equal surface potential for a conducting grain. Thus, the local electric field strength $\tilde{E}_{\rm Coul}\sim (a/\tilde{a}) E_{\rm Coul}$ is much larger than that of the main grain body, $E_{\rm Coul}\sim Z_{\rm g}e/a^2$. This means that, as $Z_{\rm g}$ increases due to hard X-ray ionization, single ions might be ejected from the tip of the substructure, or alternatively, the substructure might break apart from the main body (depending on the local tensile strength and chemical composition). Furthermore, the fractional charge of the substructure, which is proportional to $\tilde{Z_{\rm g}}/\tilde{a}^3\sim (a/\tilde{a})^2 Z_{\rm g}/a^3$, is much larger than that of the main body. Therefore, ejections of single/clustered ions from sharp substructures make it possible to efficiently remove the excessive charge from the grain, which then remains only weakly charged $Z_{\rm g}\ll Z_{\rm g,max}$ and may not be destroyed by the electrostatic stress.

Consider the physical situation that a dust grain loses $F_{\rm X}\bar{\sigma}_{\rm e}N_{\rm e}$ electrons, where $F_{\rm X}$ is the X-ray photon fluence above $\sim$$5\rm \, keV$, $\bar{\sigma}_{\rm e}\sim 10^{-22}\rm\, cm^2$ is the effective absorption cross-section per electron, and $N_{\rm e}$ is the total number of electrons in the grain. Let us suppose that the excessive positive charge is mostly carried away by the electrostatic ejection of a fraction $f_{\rm esc}$ of the mass and that the fractional charge in the ejected mass is $q_{\rm c}$ ($=$ charge number divided by the number of electrons in the ejected mass), so we write $F_{\rm X}\bar{\sigma}_{\rm e}=f_{\rm esc}q_{\rm c}$. In this scenario, grain destruction can still occur if the X-ray fluence is sufficiently high such that $f_{\rm esc}=F_{\rm X}\bar{\sigma}_{\rm e}/q_{\rm c}\sim 1$. In the limiting case of single-ion ejection, we expect $q_{\rm c}\simeq 1/Z$ ($Z$ being the atomic number) for doubly ionized ions, e.g., O$^{2+}$, Mg$^{2+}$ \citep[see Table 1 of][]{brandon1968field_evaporation}, because only the outer-shell electrons may possibly be stripped by a static electric field of a few $\rm V/\AA$. For a given $q_{\rm c}$, we see that grain destruction by ion-field emission requires an X-ray photon fluence of $F_{\rm X}\gtrsim 10^{21}\,\mr{cm}^{-2} (q_{\rm c}/0.1)^{-1}$ or isotropic equivalent energy of $E_{\rm X}(\sim 5\mr{\, keV})\gtrsim 10^{51}\mr{\,erg}\, (q_{\rm c}/0.1)^{-1} (r/1\mr{\,pc})^2$ at a distance of $r$ from the radiation source. Therefore, we conclude that irregularly shaped, conducting grains may evaporate due to ion-field emission (instead of Coulomb explosion) within a critical distance from the merger
\begin{equation}
    r_{\rm ion}\simeq 1\mr{\, pc}\, (E_{\rm X}/10^{51}\rm\, erg)^{1/2} (q_{\rm c}/0.1)^{1/2}.
\end{equation}
For typical short GRBs with isotropic equivalent jet kinetic energy $E_{\rm j}\sim\,$a few$\times10^{52}\rm\, erg$, we generally expect $E_{\rm X}(\sim 5\mr{\,keV})\lesssim 10^{51}\rm\, erg$. It requires fine-tuning for the forward shock to convert 10\% of the jet energy into X-rays in a narrow energy range between $\sim$5 and 20 keV (photons $\gtrsim20\rm\, keV$ would have been detected by \textit{Swift} BAT). Therefore, we expect
$r_{\rm ion}$ to be smaller than the thermal sublimation radius of $r_{\rm sub}\simeq 3\mr{\,pc}\, L_{\rm UV, 48}^{1/2} a_{0.1\um}^{-1/2}$, provided that $q_{\rm c}$ is close to the single-ion limit. If this is the case, the IR dust echo predicted in this work should be detectable\footnote{The referee pointed out that if IR dust echoes are observed consistent with dust destruction by sublimation only, that would argue against Coulomb explosions; on the other hand, if IR dust echoes are not seen, that might support Coulomb explosions (or some other fast mechanism) dominating dust destruction.}.

Future work on the electrostatic charge ejection process is needed to provide a reliable estimate of $q_{\rm c}$. Another uncertainty is how the grain may adjust its structure at high temperatures and under significant mass loss due to ion-field emission. Estimating the grain conductivity is beyond the scope of this work, but we point out that, at sufficiently high temperatures $T\gtrsim 1000\rm\, K$ (or $kT\gtrsim 0.1\rm\, eV$), a significant fraction of electrons on the high-energy tail will acquire an energy of $\sim 1\rm\, eV$, and these electrons will be able to conduct electricity.

\section{Summary}\label{sec:summary}
A growing number of bNS mergers will be discovered by GW detectors in the near future. Identifying their EM counterparts is an urgent task of the astronomical community. In this work, we propose a new, observable EM counterpart --- a fraction (a few to 10\%) of GW-selected bNS mergers should have bright infrared emission from the dust grains heated by the UV-optical and soft X-ray radiation from an off-axis relativistic jet. These are the mergers occurring in star-forming galaxies where the nearby environment (at distances of a few to tens of pc) has high gas density $n_{\rm H}\gtrsim 0.5\rm\, cm^{-3}$ and dust extinction $A_{\rm V}\gtrsim 0.5\rm\, mag$. For small viewing angles $\theta_{\rm obs}\lesssim 30^{\rm o}$, the dust echo emission reaches the peak flux a few months to years after the merger and should be detectable by JWST in the near- to mid-IR ($\lambda_{\rm obs}\sim2$ to $4\um$), with a detection rate of the order of once per year. We further show that for nearby sources within 150 Mpc, the spatial separation between the dust emission flux centroid and the merger site is a few to 10 mas, which may be resolvable by JWST for sufficiently high SNR ($\sim 10$) detections. Thanks to the much brighter kilonova emission, the position of the merger site can be determined to a much better precision than the dust echo, provided that the event can be localized by EM follow-ups within the first month or so. Then, astrometric measurement of the superluminal apparent motion of the dust echo directly gives the orbital inclination of the merger, which can then be combined with gravitational wave data to measure the Hubble constant \citep{hotokezaka19_hubble_constant}. We also show that dust echoes may contaminate the search for kilonova candidates from short GRBs viewed on-axis, such as the case of GRB 130603B \citep{berger13_GRB130603B_kNe_candidate, tanvir13_GRB130603B_kNe_candidate}.


\section*{Data Availability}
The data underlying this article will be shared on reasonable request to the corresponding author.

\section*{Acknowledgement}
We thank the referee, Bruce Draine, for constructive comments and Paz Beniamini for helpful discussions. WL was supported by the David and Ellen Lee Fellowship at Caltech and Lyman Spitzer, Jr. Fellowship at Princeton University. The research of CFM is supported in part by NASA grant 80NSSC20K0530. The research of KPM is supported by National Science Foundation Grant AST-1911199.

{\small
\bibliographystyle{mnras}
\bibliography{refs}
}

\appendix
\section{UV-optical Emission from External Reverse Shock}\label{sec:reverse_shock}

We calculate the afterglow emission from a relativistic jet of opening angle $\theta_{\rm j}$ and isotropic equivalent energy $E_{\rm j}$ (the actual beaming corrected energy is $\pi \theta_{\rm j}^2 E_{\rm j}$ for one jet). The following calculation follows the standard GRB afterglow theory \citep[see][]{sari99_reverse_shock, kumar15_GRB_review} and the goal is to estimate the peak luminosity and duration of the UV-optical emission from the electrons accelerated by the reverse shock, as viewed by an observer (or a dust grain) close to the axis of the jet.

\begin{figure*}
\centering
\includegraphics[width=0.45\textwidth]{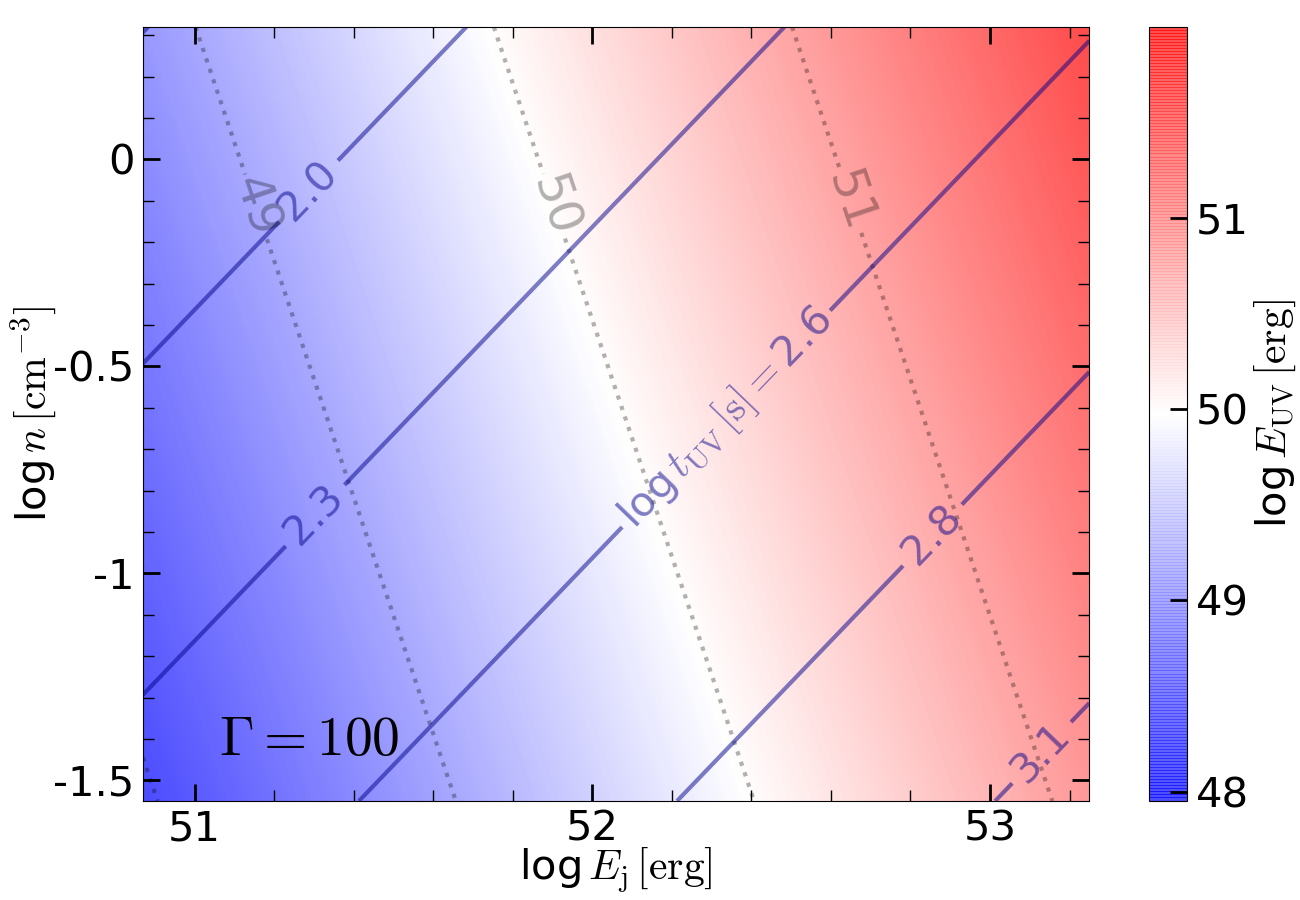}
\includegraphics[width=0.45\textwidth]{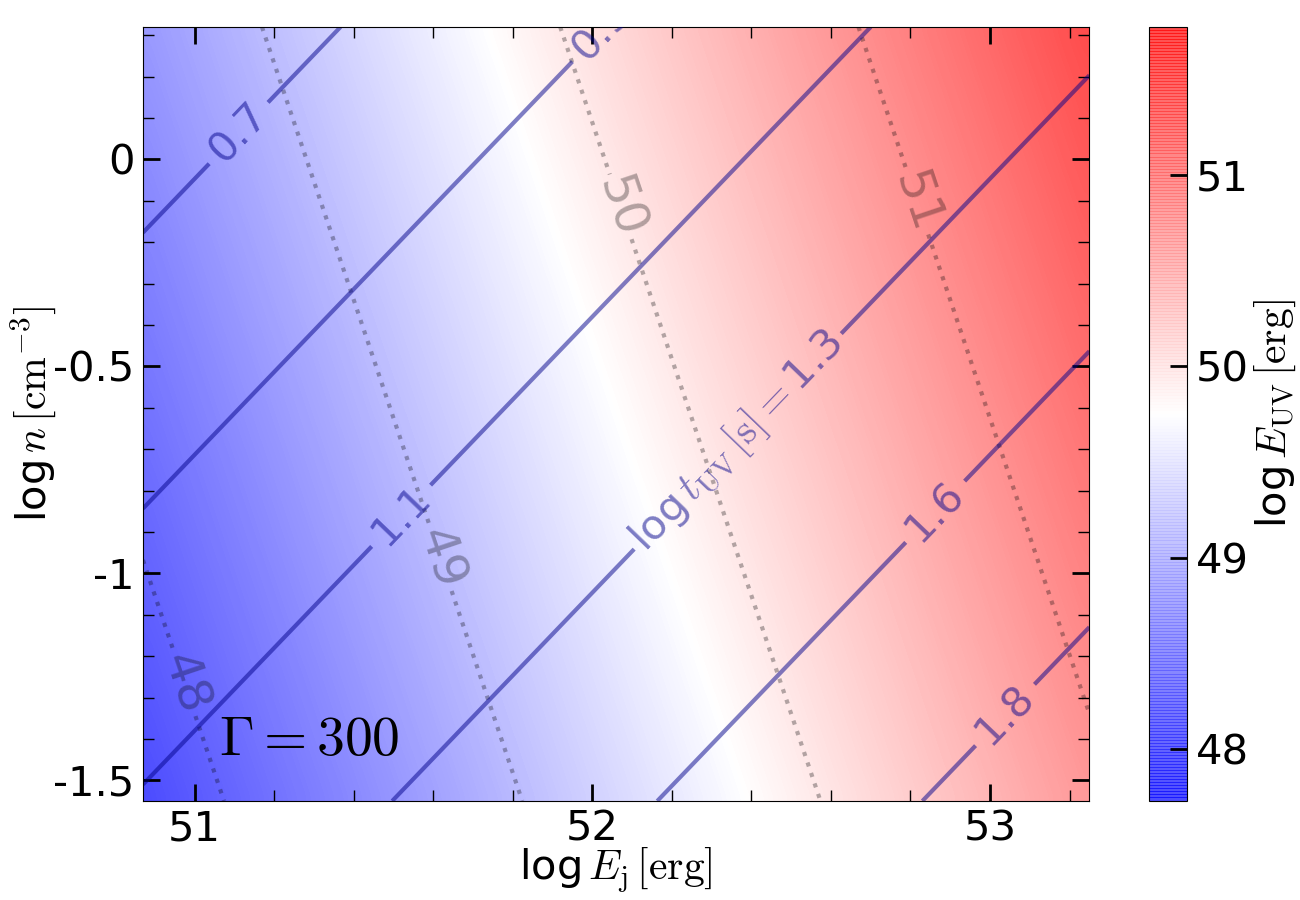}
\caption{Isotropic equivalent UV-optical radiation energy $E_{\rm UV}$ (colored image and black dotted contours) and emission duration $t_{\rm UV}$ (blue contours) for a wide range of jet energy $E_{\rm j}$ and CSM density $n$. The \textit{left panel} is for $\Gamma =100$ and the \textit{right panel} is for $\Gamma = 300$. We fix the other parameters as appropriate for typical GRBs: electron energy fraction $\epse = 0.1$, magnetic energy fraction $\epsB=0.01$, and electron power-law index $p=2.2$. We find that the UV-optical emission from the reverse shock is very diverse, with broad ranges of radiation energy and duration, $E_{\rm UV}\in (10^{48}, 10^{51.5})\rm\, erg$ and $t_{\rm UV}\in (5, 10^3)\rm\, sec$.
}
\label{fig:jet_opt}
\end{figure*}

The jet material has initial Lorentz factor $\Gamma$, and due to an order unity spread $\Delta \Gamma\sim \Gamma$, the radial extent of the jet at radius $r\gg \Gamma^2 c t_{\gamma} \simeq 3\times10^{14} \mr{cm}\, \Gamma_{2}^2 (t_{\gamma}/1\mr{\,s})$ is given by $\Delta r \sim r/2\Gamma^2$, where $t_{\gamma}$ is the duration of the prompt gamma-ray emission. The reverse shock crosses the whole jet in a deceleration time, which is given by
\begin{equation}
    t_{\rm dec} = {r_{\rm dec}\over2\Gamma^2c}  = {1\over 2\Gamma^2c} \left(3E_{\rm j}\over 4\pi \Gamma^2 n m_p c^2\right)^{1/3}\approx 195\mr{\,s}\, {E_{\rm j,52}^{1/3} \over \Gamma_{2}^{8/3} n_{-1}^{1/3}}.
\end{equation}
The speed of the reverse shock in the jet comoving frame is given by the radial extent $\Delta r' \sim r_{\rm dec}/2\Gamma$ divided by the deceleration time $t_{\rm dec}'=\Gamma t_{\rm dec}$, and this means that the reverse shock is mildly relativistic\footnote{In the case when the jet is more radially extended \citep[due to a long, lower Lorentz factor tail, e.g.,][]{uhm07_long-lived_reverse_shock}, the reverse shock may be highly relativistic at late time $t\gg t_{\rm dec}$, but the amount of energy in the tail may be small. }. Electrons are accelerated by the reverse shock to a power-law energy distribution with minimum Lorentz factor in the comoving frame of the shocked plasma
\begin{equation}
    \gm' \simeq \epse {p-2\over p-1} {m_p\over m_e}\approx 61\, \epsilon_{\rm e,-1} {3(p-2)\over p-1},
\end{equation}
where $\epse$ is the fraction of the thermal energy in the shocked plasma shared by electrons, $2<p<3$ is the power-law index ($\d N/\d \gamma\propto \gamma^{-p}$), and $m_p/m_e$ is the mass ratio between protons and electrons. The magnetic field strength in the shocked plasma is (expressed in its comoving frame)
\begin{equation}
    B' \simeq (32\pi \Gamma^2 \epsB n m_p c^2)^{1/2} \approx (1.2\mr{\,G})\,\Gamma_{2} \epsilon_{\rm B,-2}^{1/2} n_{-1}^{1/2},
\end{equation}
where $\epsB$ is the fraction of the thermal energy in the shocked plasma shared by randomly oriented magnetic fields \citep[$\epsB\sim 10^{-2}$ is favored by the bright optical peaks of some GRBs,][]{sari99_reverse_shock, kobayashi00_reverse_shock, japelj14_RS_statistics}, and we have made use of the fact that the pressure of the shocked jet is comparable to that of the shocked CSM. The characteristic synchrotron frequency corresponding to Lorentz factors of $\gm$ is given by (in the observer's frame)
\begin{equation}
    \num = \Gamma {3(\gm')^2eB'\over 4\pi m_e c} \approx (1.9\times10^{12} \mr{\,Hz})\, \Gamma_{2}^2 \epsilon_{\rm e,-1}^{2} \epsilon_{\rm B,-2}^{1/2} n_{-1}^{1/2} \left[3(p-2)\over p-1\right]^2,
\end{equation}
At observer's frequency $\nu\simeq \num$, the isotropic equivalent luminosity from synchrotron emission is given by the total number of emitting electrons $N\simeq E_{\rm j}/(\Gamma^3 m_p c^2)$ in the visible cone of opening angle $1/\Gamma$, specific power per electron $P_{\nu'} \simeq \sqrt{3}e^3B'/(m_e c^2)$ in the plasma's comoving frame, and Doppler boosting $L_\nu \simeq \Gamma^3 L_{\nu'}$, i.e.
\begin{equation}
    L_{\num} \simeq {E_{\rm j} \over m_p c^2} {\sqrt{3}e^3B\over m_e c^2} \approx (6.0\times10^{33}\mr{\,erg\,s^{-1}\, Hz^{-1}})\, E_{\rm j,52} \epsilon_{\rm B,-2}^{1/2} n_{-1}^{1/2}.
\end{equation}
The UV-optical frequency $\nu\sim 10^{15}\rm\, Hz$ is much higher than $\num$, so the flux density is given by
\begin{equation}
    L_\nu \simeq L_{\num} (\nu/\num)^{(1-p)/2},
\end{equation}
where we have ignored synchrotron cooling (unimportant for UV-optical frequencies). The shock-accelerated electrons continue to radiate for a duration of $t_{\rm UV}\simeq 2t_{\rm dec}$ until they adiabatically cool (due to expansion of the jet material), so the total energy released in the UV-optical band is given by
\begin{equation}
    E_{\rm UV}\simeq 2t_{\rm dec} \int^{\nu_{\rm max}} \d \nu L_\nu = {4t_{\rm dec}\over 3-p} \num L_{\num} (\nu_{\rm max}/\num)^{(3-p)/2},
\end{equation}
where $h\nu_{\rm max}=50\rm\, eV$ is the maximum photon energy considered (dust absorption opacity drops significantly at higher photon energies).
The results of the UV-optical radiation energy and duration are shown in Fig. \ref{fig:jet_opt}.

\section{Comparison with Other Dust Models}
\begin{figure*}
\centering
\includegraphics[width=0.48\textwidth]{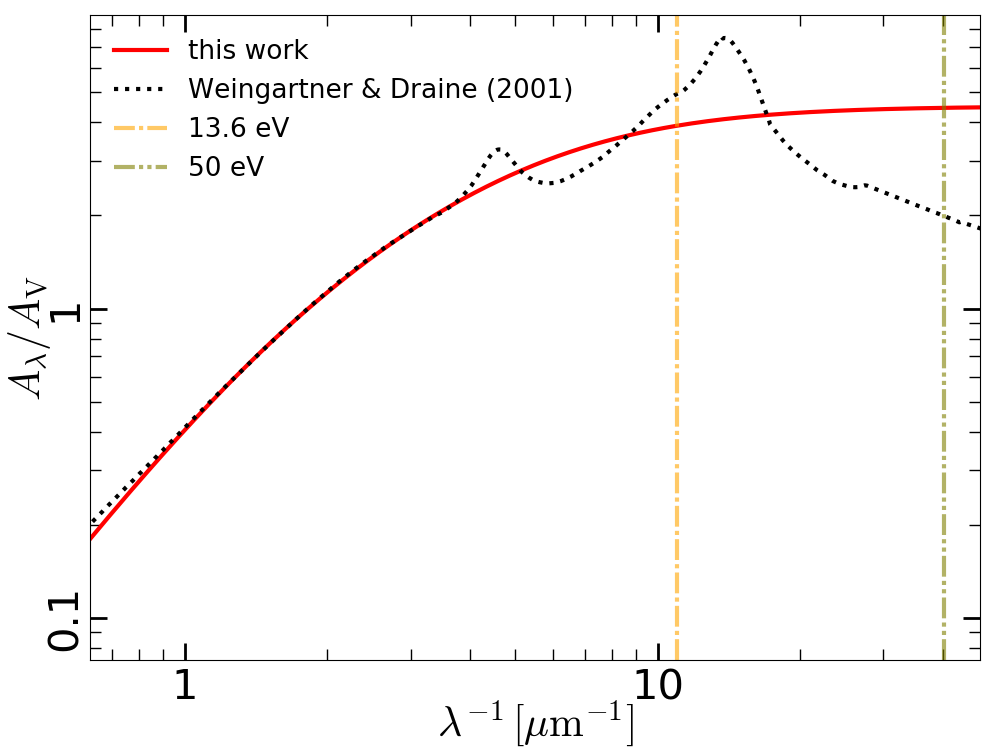}
\caption{The extinction curve given by eq. (\ref{eq:extinction_law}), for $a_{\rm min}=0.01\um$ and $a_{\rm max}=0.3\um$, as compared to the $R_{\rm V}=3.1$ curve of \citet{weingartner01_grain_size_distribution}. Our simplified model under-predicts the extinction in the $10<\lambda^{-1}<20\um^{-1}$ range but over-produces the extinction in the $20< \lambda^{-1}<40\um^{-1}$ range. The extinction at these short wavelengths is dominated by the smallest grains, which evaporate at lower temperatures up to a larger distance than bigger grains (see Fig. \ref{fig:Td}), so their contribution to the dust echo luminosity is suppressed, especially in the near-IR. Therefore, our simplified model provides a reasonably accurate description of the amount of radiation energy absorbed by dust.
}
\label{fig:extinction}
\end{figure*}

\begin{figure*}
\centering
\includegraphics[width=0.48\textwidth]{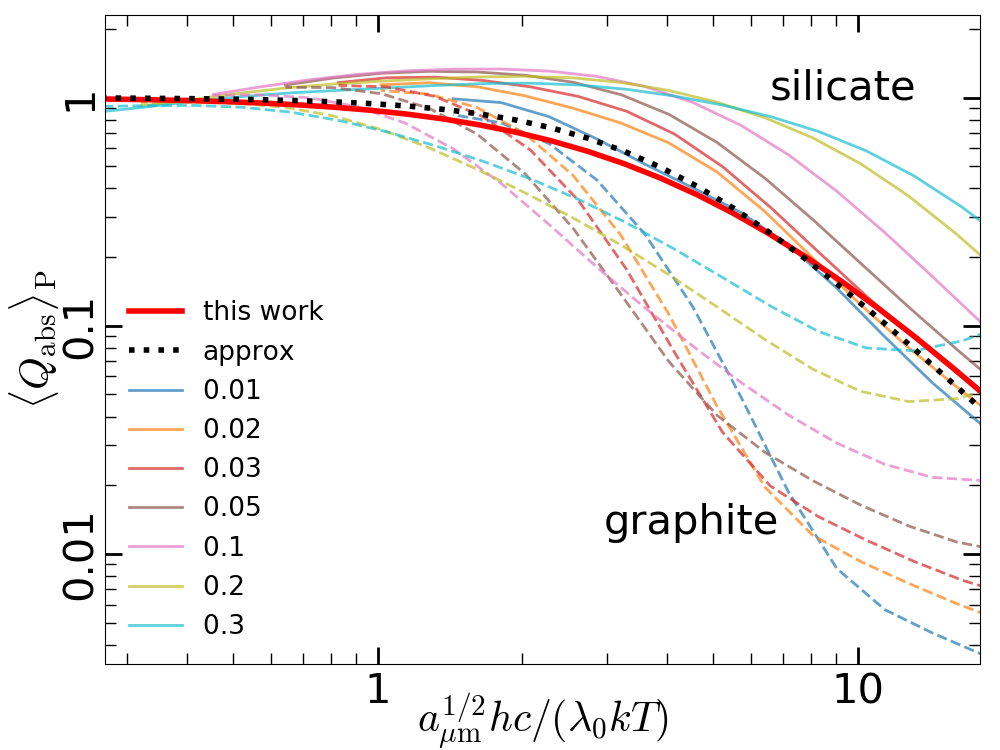}
\caption{The Planck-averaged absorption coefficient $\langle Q_{\rm abs}\rangle_{\rm P}$ as a function of grain temperature $T$. In our simplified model of eq. (\ref{eq:Qabs_lambda}), the Planck-average coefficient, as defined in eq. (\ref{eq:QabsP}), only depends on the ratio $a^{1/2}/T$, which is chosen to be the horizontal axis in this figure. The thick red line is the result of eq. (\ref{eq:QabsP}) for $\lambda_0=2\um$, and the black dashed line is the analytical approximation in eq. (\ref{eq:QabsP_approx}). The result from our model is in between the models of graphite (dashed lines) and silicate (solid lines) grain spheres by \citet{laor93_Qabs}. Their $\langle Q_{\rm abs}\rangle_{\rm P}$ has stronger dependence on the grain sizes, and different line colors correspond dust radii ranging from $a=0.01\um$ to $a=0.3\um$ as shown in the legend. The dust echo emission in the near-IR is dominated by grains near the sublimation temperature $T_{\rm sub}\sim 2000\rm\,K$, and for $0.01 < \aum < 0.3$ and $\lambda_0=2\um$, we typically have $0.3\lesssim \aum^{1/2}hc/(\lambda_0kT)\lesssim 2$, where the results from different models differ by about a factor of 2 or less (and the grain temperatures differ by about 10\% at most).
}
\label{fig:QabsP}
\end{figure*}

Our simplified dust model, consisting of the MRN size distribution (eq. \ref{eq:dnda}) and absorption/extinction coefficients (eq. \ref{eq:Qabs_lambda}), is calibrated by the extinction law and the Planck-averaged absorption coefficient taken from previous works. These two calibrations make sure that the amount of source radiation absorbed by dust is reasonably accurate and the dust grain temperatures are similar to what is expected from more sophisticated models. In Fig. \ref{fig:extinction}, we compare the extinction law given by our eq. (\ref{eq:extinction_law}) with the $R_{\rm V}=3.1$ curve of \citet{weingartner01_grain_size_distribution}, which is calibrated by the observed extinction curve \citep{cardelli89_extinction_law, fitzpatrick99_extinction_law}. In Fig. \ref{fig:QabsP}, we compare the Planck-averaged absorption coefficient given by our eq. (\ref{eq:QabsP}) with that from the model of \citet{laor93_Qabs}.

\end{document}